\documentclass[twocolumn,aps,showpacs,eqsecnum,10pt]{revtex4} 

\usepackage{latexsym}
\usepackage{amsmath}
\usepackage{amssymb}
\usepackage{epsfig}

\newcommand{\bra}[1]{\langle #1|}
\newcommand{\ket}[1]{| #1 \rangle }
\newcommand{\Bra}{\langle}

\newcommand{\E}{{\rm E}}
\newcommand{\Hs}{{\cal H}}
\newcommand{\one}{{\openone}}

\newcommand{\nstav}{{\rm \alpha_{?}}}
\newcommand{\na}{n_A}
\newcommand{\nb}{n_B}
\newcommand{\nc}{n_C}
\newcommand{\nn}{n_N}
\newcommand{\complex}{{\mathbb C}}

\newcommand{\pui}[3]{P_{#1}(\ket{#2},\ket{#3})}
\newcommand{\mpui}[1]{\overline{P_{#1}}}
\newcommand{\puim}{P(\ket{\alpha_1},\ldots,\ket{\alpha_M})}

\begin{document}
\title{Unambiguous identification of coherent states II: Multiple resources}
\author{Michal Sedl\'ak$^{1,3}$,
M\'ario Ziman$^{1,2}$,
Vladim\'\i r Bu\v zek$^{1,3}$, and Mark Hillery$^{4}$}
\address{
$^{1}$Research Center for Quantum Information, Slovak Academy of Sciences, D\'ubravsk\'a cesta 9, 845 11 Bratislava, Slovakia \\
$^{2}$Faculty of Informatics, Masaryk University, Botanick\'a 68a, 602 00 Brno, Czech Republic\\
$^{3}${\em Quniverse}, L{\'\i}\v{s}\v{c}ie \'{u}dolie 116, 841 04 Bratislava, Slovakia\\
$^{4}$Department of Physics and Astronomy, Hunter College
of the City University of New York, 695 Park Avenue, New York, NY 10021, USA
}
\begin{abstract}
We consider unambiguous identification of coherent states of electromagnetic field. 
In particular, we study possible generalizations of an optical setup proposed in M. Sedl\'{a}k {\it et al.}, 
Phys. Rev. A {\bf 76}, 022326 (2007). We show how the unambiguous identification of coherent states can be performed in a general case when multiple copies of unknown and the reference states are available.  We also investigate whether reference states after the measurement can
be ``recovered'' and further used for subsequent unambiguous identification tasks.  We show that in spite of the fact that the recovered reference states are disturbed by measurements they can be repeatedly used for unambiguous identifications. We analyze the role of various imperfections in preparation of the unknown and the reference coherent states on the performance of our unambiguous identification setup.
\end{abstract}

\pacs{03.67.Lx,02.50.Ga}
\maketitle

\section{Introduction}
Discrimination of quantum states is a challenging task which dramatically differs from discrimination of states of a classical system. The main difference between quantum and classical discriminating problems is that single copies of {\em nonorthogonal} quantum states cannot be distinguished perfectly. There exist several strategies how to approach the problem of quantum-state discrimination which are related to different choices of a figure of merit that defines what is considered to be the best measurement. In particular, the strategy that minimizes  a mean probability of error is called the {\em min-error} approach \cite{helstrom}. Another equally well motivated strategy is the so-called  unambiguous discrimination \cite{ivanovic,dieks,peres}. In this strategy errors in conclusions are not permitted, which implies that also inconclusive results are acceptable. Nevertheless, whenever we obtain a conclusive result we know for sure that it is correct. Various  discrimination tasks may differ by {\em a prior}  information about the quantum system we are measuring. The discrimination among two known pure states represents a situation, where we have the maximum  knowledge about the possible preparations of the system.
On the other hand one can consider a situation when we have no classical information available about the preparation.
Specifically, instead of being given only one unknown quantum system, we are given additional quantum systems, denoted as reference states, that represent possible preparations. Our task is to determine unambiguously with which reference states the unknown quantum system match.
This problem is coined as the Unambiguous Identification (UI) and can be used, for example, in a communication of parties that do not share common reference frame. In such case reference states can be seen as an ``alphabet'' that is always send along with the classical information carried by the unknown state. If parties are communicating over medium unstable in time (e.g. optical fiber, air,$\ldots$) then the UI overcomes a standard need for a re-calibration of communication channel induced by the changes in the medium.

In our previous paper \cite{ui1} (to which we will refer as to Paper I) we showed that the UI for modes of an electromagnetic field can be efficiently performed if we known that states to be discriminated are coherent states with unknown amplitudes. We shown that this prior knowledge (or, a reduction of a set of considered states of an electromagnetic mode) significantly improves the success of a discrimination process. Furthermore, we proposed a simple optical setup consisting of a set of beam splitters and photodetectors for implementation of optimal UI measurement. This setup was also recently experimentally realized by L. Bart\r u\v skov\'a {\em et al.} \cite{dusek}.

In the present paper we study possible generalizations of an optical setup proposed in Paper I and we investigate modifications and generalizations of  the unambiguous identification  of coherent states. More specifically, in Sec. \ref{genstrat1} we show how the UI of coherent states can be performed in a general case when multiple copies of unknown and reference states are available. This investigation is motivated by the observation that with the increase of the number of identically prepared particles we can better identify the preparator, so we are "closer" to a classical domain. In Sec. \ref{recov1} we investigate the possibilities for a recovery of reference states after the measurement is performed. This recovery process might seem to be prohibited by the rules of quantum mechanics (due to irreversible disturbance of a quantum state by a measurement). Nevertheless, we show that in spite of the fact that the recovered reference states are ``degraded'' (disturbed), nevertheless they can be used in the next round of UI - this is true under the condition that an undisturbed (new) copy of an unknown state is provided. Such a scenario can be seen as a repeated search in a quantum database, where the data, i.e. the reference states, degrade with their repeated use. In Sec IV we investigate the influence of various imperfections and a noise on the performance of our UI setups. Our findings are summarized in Sec. V while the Appendix contains some technical details of our calculations and proofs.

\section{General UI measurement strategy with beam splitters}
\label{genstrat1}

We start our investigation of  unambiguous identification tasks with posing the problem within a sufficiently wide framework. Specifically, we shall consider a set of modes of quantum electromagnetic field (linear harmonic oscillators)
each of which with a semi-infinite-dimensional Hilbert space $\Hs_\infty$. In addition, we shall consider that each of the mode  is prepared in a coherent state of a specific amplitude. All together we will consider a set of $M+1$ groups of electromagnetic modes.
The group  $A$ of $\na$ modes $A_1,\ldots,A_{\na}$ carries $\na$ copies of an unknown coherent state $\ket{\nstav}$. Each of the remaining $M$ groups $B,C,D,\ldots$ contains $\nb, \nc, n_D, \ldots$ copies of the reference states $\ket{\alpha_1}, \ket{\alpha_2}, \ket{\alpha_3},\ldots,\ket{\alpha_M}$, respectively. Moreover, we are guaranteed that the unknown state $\ket{\nstav}$ is the same as one of the reference states $\ket{\alpha_k}$ $k=1,\ldots,M$. Our task is to find out unambiguously with which reference state the unknown state matches.
Thus, in general, the discrimination problem corresponds to a selection among one of $M$ possible types of states:
\begin{eqnarray}
\ket{\Psi_i}_{ABC\ldots}&\equiv&\ket{\alpha_i}^{\otimes\na}_{A}\otimes\ket{\alpha_1}^{\otimes\nb}_{B}\otimes\ket{\alpha_2}^{\otimes\nc}_{C}\otimes\ldots \nonumber\\
& & \ldots\otimes \ket{\alpha_1}^{\otimes\nn}_{N},
\label{stavy1}
\end{eqnarray}
with $ i=1,2,\ldots,M $.
The most general measurement strategy unambiguously discriminating among these $M$ possibilities can be described by a positive operator value measure (POVM) consisting of $M+1$ elements $E_i$ with $i=0,1,\ldots,M$. The element $\E_i$ corresponds to a correct identification of $\ket{\Psi_i}$  while $\E_0$ corresponds to an inconclusive result, i.e. the failure of the measurement.
These elements must obey ``no-error'' conditions [Eq. (\ref{stavy2})] and they have to constitute a proper POVM [Eq. (\ref{povm})]:
\begin{eqnarray}
\forall i\neq j \quad Tr[\E_i\rho_j]=0; \quad \rho_i=\ket{\Psi_i}\bra{\Psi_i} \label{stavy2}; \\
\E_i\geq0,\E_0\geq0; \quad \E_0+\sum^{M}_{i=1}\E_i=\one. \label{povm}
\end{eqnarray}
We assume that the states of the type $\ket{\Psi_i}$ appear with an equal prior probability $\eta_i=1/M$. The performance of this UI measurement can be quantified by a probability of identification for a particular choice of reference states
\begin{eqnarray}
\puim=\sum^{M}_{i=1}\eta_i Tr[\E_i\rho_i]\label{pidentify}\, .
\end{eqnarray}
However,  a more adequate figure of merit is its average value, i.e.
\begin{eqnarray}
\mpui{}=\int_{\complex^M} \puim \chi(\alpha_1,\ldots,\alpha_M)\ d\alpha_1\ldots d\alpha_M , \nonumber \\
\label{meanpi}
\end{eqnarray}
where $\chi(\alpha_1,\ldots,\alpha_M)$ is the probability distribution describing our knowledge about the choice of reference states.
The optimality of a UI measurement is defined with respect to the average performance $\mpui{}$. However, we will see that in most of the optical setups we propose it suffices to optimize $\puim$, because the optimal value of the transmittivities in the setup does not depend on specific reference states. In Eq. (\ref{meanpi}) we integrate over multiple infinite (complex) planes of complex amplitudes.
 Unfortunately, a uniform distribution on an infinite plane can not be properly defined. Thus, $\chi(\alpha_1,\ldots,\alpha_M)$ can not be uniform, but instead should be ``regularized'', i.e. it should satisfy some reasonable physical requirements. For example, the probability of having reference states with very big amplitudes, i.e. of very high energy, should be vanishing. In the present paper we will calculate $\puim$, because most of the features that the averaged probability $\mpui{}$ will have are already apparent in $\puim$.

In Paper I we focused our attention mainly on the UI problem with a single copy of an unknown state and a single copy of each of two reference states ($M=2$,$\na=\nb=\nc=1$). We proposed a simple efficient UI measurement utilizing three beam splitters and two photodetectors.
 The whole setup is supposed to operate as follows:
 The unknown state $\ket{\nstav}$ is split by the first $50/50$ beam splitter. As a result we obtain  two equally ``diluted''
copies of the unknown state \footnote{Here we refer to ``diluted'' states in a sense that the original (unknown) information that has been encoded in the amplitude $\nstav$ of a coherent state of a single mode is distributed into two modes each in a coherent state with the amplitude $\nstav/\sqrt{2}$. Due to the unitarity of the process no information was lost.} described by a vector  $\ket{\frac{1}{\sqrt{2}}\nstav}\otimes\ket{\frac{1}{\sqrt{2}}\nstav}$.
 Each of these ``copies'' $\ket{\frac{1}{\sqrt{2}}\nstav}$ is, together with one of the reference states, fed into the second (respectively the third) unbalanced beam splitter. The second (respectively the third) beam splitter performs the comparison of the diluted unknown state $\ket{\frac{1}{\sqrt{2}}\nstav}$ with the first (respectively the second) reference state. Detection of photons at the output modes of these quantum-state comparison measurement setups \cite{jex,andersson,comparison} unambiguously indicates that $\ket{\nstav}$ differs from $\ket{\alpha_1}$ or $\ket{\alpha_2}$. This enables us to conclude that $\ket{\nstav}=\ket{\alpha_2}$ or $\ket{\nstav}=\ket{\alpha_1}$, respectively. We show in Appendix \ref{app1}, that this measurement is optimal if we restrict ourselves to discrimination of coherent states with the use of linear optical elements, number resolving photodetectors.

Naturally, coherent states encode complex numbers. From this point of view the state $\ket{\frac{1}{\sqrt{2}}\nstav}$ carries formally the whole information about the complex amplitude.
 This is due to the fact that we know the factor $\lambda=1/\sqrt{2}$ by which $\nstav$ is rescaled. If the complex amplitude $\nstav$ is encoded in the state $\ket{\lambda\nstav}$ then for $0\leq\lambda<1$ we will speak about  a ``diluted'' unknown state while the case $\lambda>1$ will be referred to as a ``concentrated'' unknown state $\ket{\nstav}$. These terms come from  the fact that the ``diluted'' state can be obtained by mixing a coherent state and a vacuum at an input of a beam splitter. As a result of the beam splitter transformation two modes at the output of the beam splitter are in the diluted states.
On the contrary, the ``concentrated'' state can be prepared by  launching two copies of the same coherent state into the beam splitter.
As a result we obtain one of the output modes in the ``concentrated'' state while the second mode in the vacuum state. Using a sequence of beamsplitters and corresponding resources one can prepare ``diluted' or ``concentrated'' states with arbitrary value of the scaling factor $\lambda$.
\begin{figure}
\begin{center}
\includegraphics[width=7cm]{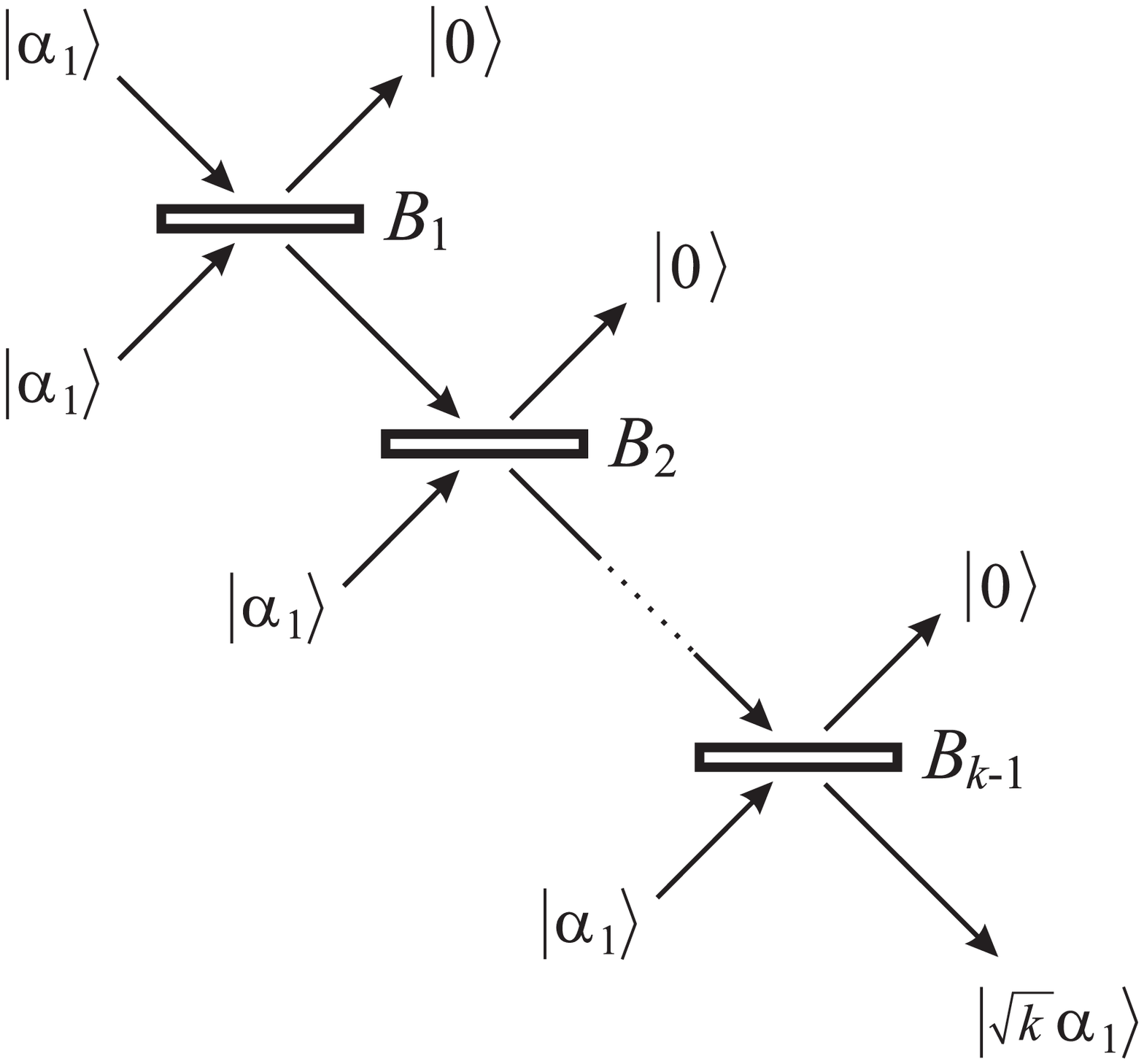}
\caption{The beamsplitter setup designed for constructive interference of the same input coherent states.
Out of $k$ copies of a coherent state $\vert \alpha\rangle$ we obtain at the output of a sequence of $k-1$ beamspitters one mode in the coherent state $\vert \sqrt{2}\alpha\rangle$ and $k-1$ modes in a vacuum state $|vert 0 \rangle$.
}
\label{collector}
\end{center}
\end{figure}
Actually, preparation of ``concentrated'' states is the main idea we will employ in our investigation of the UI measurement with multiple copies of unknown and reference states. At the beginning of the UI measurement we will, for each kind of a state, concentrate the information encoded in its $k$ copies into a single quantum system. This can be done by a sequence of $k-1$ beam splitters (see Fig.~\ref{collector}) with transmitivities chosen so that the input state $\ket{\beta}^{\otimes k}$ constructively interferes to produce the state $\ket{\sqrt{k}\beta}\otimes\ket{0}^{\otimes k-1}$. More details about this transformation can be found in Sec. III.A of \cite{comparison}. The result of these preliminary transformations is a mapping of possible types of states $\ket{\Psi_i}$ into states $\ket{\sqrt{\na}\alpha_i}_{A_1}\otimes\ket{\sqrt{\nb}\alpha_1}_{B_1}\otimes\ket{\sqrt{\nc}\alpha_2}_{C_1}\otimes\ldots\otimes\ket{0}^{t}$, where $t=\na-1+\nb-1+\ldots+n_M-1$. As a next step we will use the setup proposed in Sec. III of Paper I for a single copy of the unknown state and  single copies of $M$ reference states. Of course, as we will see below the transmitivities of all beam splitters in the setup must be modified according to the number of copies of the unknown and the reference states we are given.

\subsection{Two types of reference states}
\label{tworstates}

\begin{figure}
\begin{center}
\includegraphics[width=7cm]{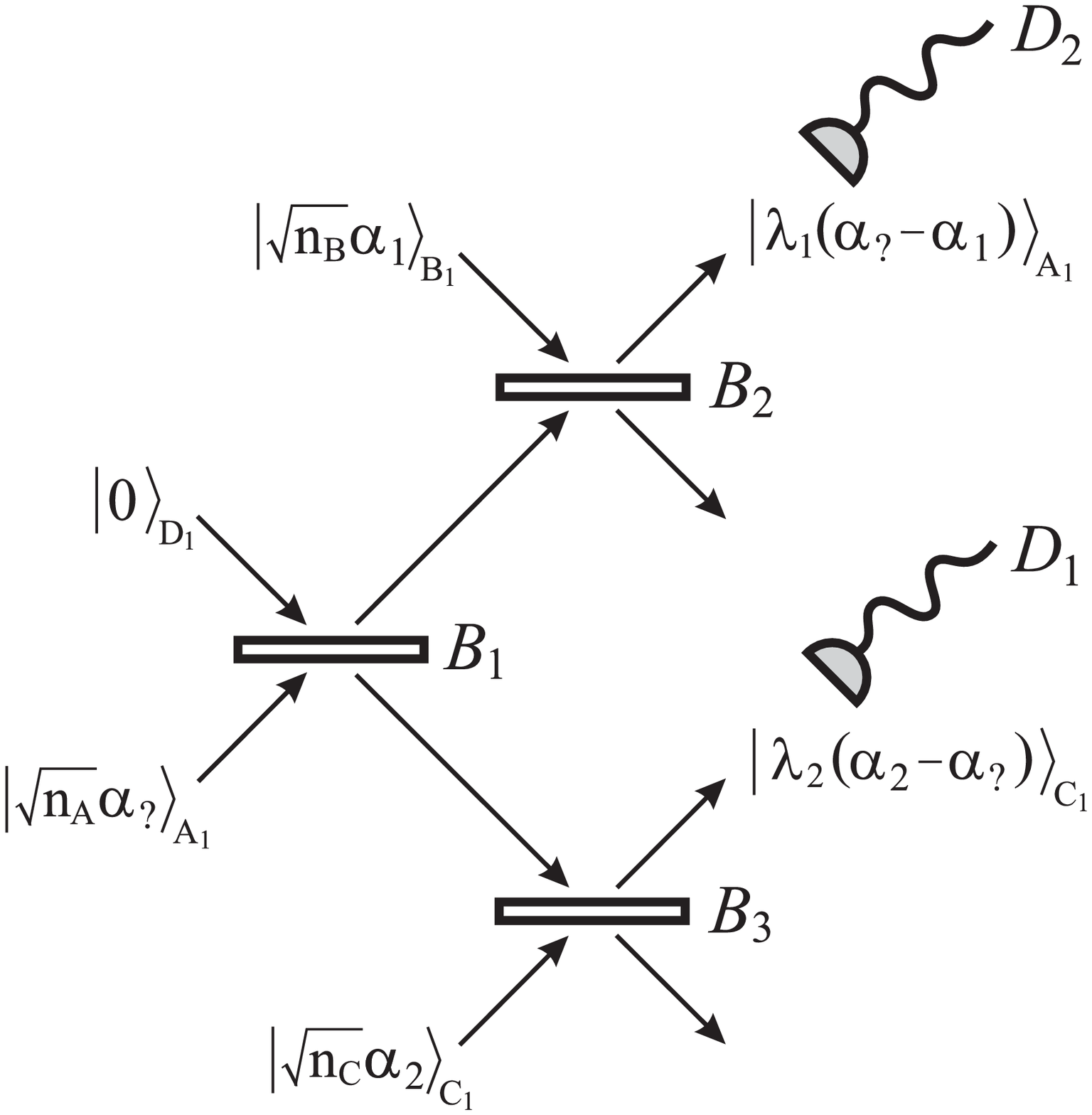}
\caption{The beamsplitter setup designed for an unambiguous identification of multiple copies of
two types of coherent states.
}
\label{multi1}
\end{center}
\end{figure}
The unambiguous identification of two types of coherent reference states is the first natural step in generalizing the scenario with single copies of unknown and reference states investigated in Paper I. In this section we consider $M=2$ and $\na, \nb, \nc$ are arbitrary. The above mentioned idea of ``concentration'' of quantum information implies that we first feed all provided copies of the unknown state into $\na-1$ beam splitters to obtain the first state $\ket{\sqrt{\na}\nstav}$ in the mode $A_1$ (for brevity later called only $A$). The other modes $A_2 \ldots A_{\na}$ end up in a vacuum state, therefore we will not consider them further. Similarly, $\nb-1$ (respectively, $\nc-1$) beam splitters are used to prepare the state $\ket{\sqrt{\nb}\alpha_1}$ (respectively, $\ket{\sqrt{\nc}\alpha_2}$) in the modes $B_1$ ($C_1$). Next, we feed these concentrated states into essentially the same scheme as in Paper I (see Fig.~\ref{multi1}). Thus, altogether we are going to use $\na+\nb+\nc$ beam-splitters. The analysis of the setup presented in Fig.~\ref{multi1} is analogous to Paper I, therefore we comment on it only briefly.

A beamsplitter transforms two input modes prepared in coherent states $\ket{\alpha}$ and $\ket{\beta}$, respectively, as
$\ket{\alpha}\otimes\ket{\beta}\mapsto\ket{\sqrt{T}\alpha+\sqrt{R}\beta}\otimes\ket{-\sqrt{R}\alpha+\sqrt{T}\beta}$
where $R,T$ stands for a reflectivity and a transmittivity coefficients
of the beamsplitter. The setup in Fig.~\ref{multi1} employs one additional input mode $D$ initially
prepared in a vacuum state, i.e. the state vector describing four input modes reads
\begin{eqnarray*}
\ket{\Phi_{in}}=\ket{\sqrt{\na}\nstav}_A\otimes\ket{\sqrt{\nb}\alpha_1}_B
\otimes \ket{\sqrt{\nc}\alpha_2}_C \otimes \ket{0}_D\, ,
\end{eqnarray*}
where $\nstav$ is guaranteed to be either $\alpha_1$ or $\alpha_2$.
The action of the three beamsplitters in the setup is described by
a unitary transformation
\begin{eqnarray}
\nonumber
\ket{\Phi_{in}}\mapsto \ket{\Phi_{out}}=(U^{(2)}_{AB} \otimes U^{(3)}_{CD})
(U^{(1)}_{AD}\otimes I_{BC})\ket{\Phi_{in}} \, ,
\end{eqnarray}
where $U^{(j)}_{XY}$ is associated with the $j$-th
beamsplitter $B_j$ acting on the modes $X$ and $Y$.
Since beamsplitters do not entangle coherent states it follows
that the output state $\ket{\Phi_{out}}$ remains factorized.
In the first step the beamsplitter $B_1$ with transmittivity $T_1$
prepares two ``diluted'' copies of the state $\ket{\sqrt{\na}\nstav}$, i.e.
\begin{eqnarray}
\ket{0}_D\otimes\ket{\sqrt{\na}\nstav}_A \mapsto \ket{\sqrt{R_1 \na}\nstav}_D\otimes\ket{\sqrt{T_1 \na}\nstav}_A\, .
\end{eqnarray}
In the second step the beamsplitters $B_2,B_3$ perform the transformation
such that the output state reads
\begin{eqnarray}
\ket{\Phi_{out}}=\ket{out}_A\otimes\ket{out}_B\otimes\ket{out}_C
\otimes\ket{out}_D  \, ,
\end{eqnarray}
with
\begin{eqnarray}
\nonumber
\ket{out}_A &=& \ket{-\sqrt{R_2\nb}\alpha_1+\sqrt{T_2 T_1 \na}\nstav}_A\, , \\
\nonumber
\ket{out}_B &=& \ket{\sqrt{T_2\nb}\alpha_1+\sqrt{R_2 T_1\na}\nstav}_B\, , \\
\nonumber
\ket{out}_C &=& \ket{-\sqrt{R_3 R_1 \na}\nstav+\sqrt{T_3 \nc}\alpha_2}_C \, ,\\
\nonumber
\ket{out}_D &=& \ket{\sqrt{T_3 R_1 \na}\nstav+\sqrt{R_3 \nc}\alpha_2}_D\, .
\end{eqnarray}
A crucial observation is that the parameters $T_j,R_j=1-T_j$ can be adjusted
so that either the mode $A$, or the mode $C$, ends up in a vacuum state
providing that $\alpha_?=\alpha_1$, or $\alpha_?=\alpha_2$, respectively.
In particular, setting the transmittivities to
\begin{eqnarray}
T_2=\frac{1}{1+\frac{\na}{\nb} T_1}\, ; \quad
T_3=\frac{1-T_1}{\frac{\nc}{\na}+1-T_1}\, ,
\label{t2set}
\end{eqnarray}
we find
\begin{eqnarray}
\nonumber\ket{out}_A&=&\ket{\sqrt{R_2\nb}(\nstav-\alpha_1)}_A\, ;\\
\nonumber\ket{out}_B&=&\ket{\sqrt{T_2 \nb}\alpha_1+\sqrt{R_2 T_1\na}\nstav}_B\, ;\\
\nonumber\ket{out}_C&= &\ket{\sqrt{T_3\nc}(\alpha_2-\nstav)}_C\, ;\\
\ket{out}_D&= &\ket{\sqrt{T_3 R_1 \na}\nstav+\sqrt{R_3\nc}\alpha_2}_D
\, . \label{bs23stav}
\end{eqnarray}

Finally, we perform photodetection in output the modes $A$ and $C$
by photodetectors $D_2$ and $D_1$, respectively. By detecting a photon in one of the two modes we can
unambiguously identify the unknown state. In particular, for these two modes we have
\begin{eqnarray}
\nstav=\alpha_1 \leftrightarrow \ket{0}_A
\otimes\ket{\sqrt{T_3 \nc}(\alpha_2-\alpha_1)}_C \, ;\nonumber\\
\nstav=\alpha_2 \leftrightarrow \ket{\sqrt{R_2 \nb}(\alpha_2-\alpha_1)}_A
\otimes\ket{0}_C \, .
\label{identifstavy}
\end{eqnarray}
We note that due to the fact that at least one of the modes is in a vacuum state
both detectors cannot ``click'' (detect the photons)
at the same time. Therefore, in each single run of the experiment only three situations
can happen:\newline
{\it i)} none of the detectors click,\newline
{\it ii)} only the detector $D_1$ clicks,\newline
{\it iii)} only the detector $D_2$ clicks.\newline
If only the detector $D_1$ clicks that following Eqs. (\ref{identifstavy})
we unambiguously conclude that $\nstav=\alpha_1$. Similarly, if only
the detector $D_2$ clicks we unambiguously conclude
that $\nstav=\alpha_2$. If none of the detectors click we cannot
determine which mode was not in the vacuum state and therefore this situation represents
an inconclusive result.

If $\nstav=\alpha_1$, then the probability of a correct identification
is given as the probability of detecting at least one photon in the mode $C$
\begin{equation}
P_1=1-|\Bra{0}\ket{\sqrt{T_3\nc}(\alpha_2-\alpha_1)}|^2
=1-e^{-\frac{\nc \na(1-T_1)}{\nc+\na(1-T_1)}|\alpha_1-\alpha_2|^2}\, .
\end{equation}
Analogously, in the case $\nstav=\alpha_2$ the probability of a correct
identification reads
\begin{equation}
P_2=1-|\Bra{0}\ket{\sqrt{R_2 \nb}(\alpha_2-\alpha_1)}|^2
=1-e^{-\frac{\nb \na T_1}{\nb+\na T_1}|\alpha_1-\alpha_2|^2}\, .
\end{equation}
Thus the total probability of the identification of reference states
$\ket{\alpha_1}$ and $\ket{\alpha_2}$ is equal to
\begin{eqnarray}
\pui{}{\alpha_1}{\alpha_2}=\eta_1 P_1+\eta_2 P_2=\frac{1}{2}(P_1+P_2)\, .
\label{pibsstavy}
\end{eqnarray}

Next we will optimize the performance of the setup by
choosing an appropriate value of transmittivity $T_1$. The definition of the uniform
distribution on the set of coherent states is problematic,
therefore we first focus our attention on the probability of identification for
a particular choice of reference states
$\ket{\alpha_1}$ and $\ket{\alpha_2}$, respectively, expressed by Eq.~(\ref{pibsstavy}).

The investigation of the first derivative $\frac{\partial \pui{}{\alpha_1}{\alpha_2}}{\partial T_1}$ reveals that the optimal choice of $T_1$ does not depend on the reference states $\ket{\alpha_1}$, $\ket{\alpha_2}$ only if $\nb=\nc$. As one expects, because of symmetry arguments,  $T_1$ is optimally set to $1/2$ if $\nb=\nc$. In such a case, $\pui{}{\alpha_1}{\alpha_2}$ can be simplified to take the following form:
\begin{eqnarray}
\pui{}{\alpha_1}{\alpha_2}=1-e^{-\frac{\na \nb}{\na+2\nb}|\alpha_1-\alpha_2|^2}\, .
\label{puibs}
\end{eqnarray}
Let us note that if $\nb \neq \nc$ then there exists a prior probability $\eta_1=1-\eta_2$ for which the optimal choice of $T_1$ does not depend on the reference states. However, as already mentioned, we focus on the $\eta_i=1/M$ case and we will assume that we are given the same number of copies of each reference state.

\subsubsection{Trade-off of resources}

The number of copies of an unknown state or of a reference state we have can be seen as a measure of some resource. From this point of view an interesting question immediately arises. Which type of resource is more useful for an unambiguous identification of coherent states? Are unknown states more useful than reference states or vice versa? To answer these questions we consider the following situation. Imagine we will get altogether $N$ quantum systems (modes of electromagnetic field) but we have a liberty to specify whether the specific mode is prepared in
in the unknown state or in one of the two reference states. Thus, if we ask for $\na$ copies of the unknown state we will obtain $\nb=\nc=(N-\na)/2$ copies per a reference state. Let us for simplicity assume that $N$ and $\na$ have the same parity. The probability of identification for a reference states $\ket{\alpha_1}$, $\ket{\alpha_2}$  then reads
\begin{eqnarray}
\pui{}{\alpha_1}{\alpha_2}=1-e^{-\frac{\na (N-\na)}{2N}|\alpha_1-\alpha_2|^2}
\label{czdrojov1}
\end{eqnarray}
and it is maximized for $\na=\lfloor N/2\rfloor$, because the terms in the exponent are nonnegative. Hence, from the point of view of the resources, it is optimal to ask for a preparation of $\lfloor N/2\rfloor$ unknown states and the equal number of copies per a reference state (specifically, $\lfloor N/4\rfloor$).

\subsubsection{Infinite number of copies of reference states}

Unambiguous identification is a discrimination task in which we have very limited prior knowledge about the possible preparations of the quantum system we are given. The amount of information about the possible preparations is essentially given by the number of copies of the reference states we obtain. In the limit of infinite number of them the preparation of reference states become known to us and thus the UI is becoming equivalent to discrimination among {\em known} states. The unambiguous discrimination among pair of known pure states (for equal prior probabilities) was solved by Ivanovic, Dieks and Peres \cite{ivanovic,dieks,peres} in 1987. Their optimal measurement succeeds with a probability $1-|\Bra{\varphi_1}\ket{\varphi_2}|$, where $\ket{\varphi_1}$, $\ket{\varphi_2}$ are the known states in which the system can be prepared. In what follows we will show that in the aforementioned limit ($M=2, \nb=\nc\rightarrow\infty$) our beam-splitter setup achieves the same optimal performance. In order to prove this we have to evaluate the limit of Eq. (\ref{puibs}):
\begin{eqnarray}
P(\ket{\alpha_1},\ket{\alpha_2}, \nb&=&\nc\rightarrow \infty) \nonumber\\
&=&\lim_{\nb\rightarrow\infty} 1-e^{-\frac{\na \nb}{\na+2\nb}|\alpha_1-\alpha_2|^2} \nonumber\\
&=&1-e^{-\frac{\na}{2}|\alpha_1-\alpha_2|^2}\nonumber\\
&=&1-|\Bra{\alpha_1}\ket{\alpha_2}|^{\na}.
\label{puibslim}
\end{eqnarray}
In the last equality we have used the expression for the modulus of the overlap of the two coherent states $|\Bra{\alpha_1}\ket{\alpha_2}|^2=e^{-|\alpha_1-\alpha_2|^2}$. In the limit $\nb=\nc\rightarrow \infty$ the two known states that could be unambiguously discriminated by the Ivanovic-Dieks-Peres measurement are $\ket{\varphi_1}=\ket{\alpha_1}^{\otimes \na}$, $\ket{\varphi_2}=\ket{\alpha_2}^{\otimes \na}$. Thus, we see that Eq. (\ref{puibslim}) is equal to $1-|\Bra{\varphi_1}\ket{\varphi_2}|$ and so our beam-splitter setup performs optimally in this limit.
Let us note that for $\na=1$ our setup is in this limit equivalent to the setup proposed by K. Banaszek \cite{banaszek} for unambiguous discrimination between a pair of known coherent states. For $\nb=\nc\rightarrow \infty$ our $T_2\rightarrow 1$, $T_3\rightarrow 0$, i.e. the ``concentrated'' reference states are nearly reflected, which induces a displacement of the ``diluted'' unknown state $\ket{\frac{1}{\sqrt{2}}\nstav}$. In the same way K. Banaszek uses very unbalanced beam-splitters to cause the displacement of the outputs of the beam-splitter.

For the limiting case $M=2$, $\na=\nb=\nc\rightarrow\infty$ it is natural to expect a classical behavior, i.e. a unit probability of identification. For unequal reference states this result is easily obtained by taking the limit of Eq. (\ref{puibs}).

\subsubsection{Weak implementation of UI measurement}
\label{weak1}

Let us consider a basic version of UI of coherent states ($M=2$, $\na=\nb=\nc=1$). We will describe a measurement, which in the case of success,  leaves all the input states nearly unperturbed and achieves the same probability of identification as the original setup from Paper I. The measurement procedure goes as follows. We first equally split each of our resource states into $N$ parts. Thus, we have $N$ copies of states $\ket{\frac{1}{\sqrt{N}}\nstav}, \ket{\frac{1}{\sqrt{N}}\alpha_1}, \ket{\frac{1}{\sqrt{N}}\alpha_2}$. We use the beam-splitter setup from Paper I for each of these $N$ triples. The UI measurement performed on the first triple will succeed with probability $1-e^{-\frac{1}{3}|\frac{1}{\sqrt{N}}\alpha_1-\frac{1}{\sqrt{N}}\alpha_2|^2}=1-e^{-\frac{1}{3N}|\alpha_1-\alpha_2|^2}$. If we find $\nstav=\alpha_1$ we can combine the unmeasured $3N-3$ modes into states $\ket{\sqrt{\frac{2N-2}{N}}\alpha_1}, \ket{\sqrt{\frac{N-1}{N}}\alpha_2}$. For $\nstav=\alpha_2$ we operate analogously obtaining $\ket{\sqrt{\frac{N-1}{N}}\alpha_1}, \ket{\sqrt{\frac{2N-2}{N}}\alpha_2}$. If UI measurement of the first triples fails we continue by measuring the other triples until we find a conclusive outcome or use all the triples. In case of k-th triple leading to conclusive result we concentrate the remaining resources to obtain states $\ket{\sqrt{\frac{2(N-k)}{N}}\alpha_1}, \ket{\sqrt{\frac{N-k}{N}}\alpha_2}$ or $\ket{\sqrt{\frac{N-k}{N}}\alpha_1}, \ket{\sqrt{\frac{2(N-k)}{N}}\alpha_2}$ depending on $\nstav$ being $\alpha_1$ or $\alpha_2$. We do not get a conclusive result only if the measurements of all $N$ triples yield inconclusive results. Hence, the overall probability of successful identification of the unknown states is $1-(e^{-\frac{1}{3N}|\alpha_1-\alpha_2|^2})^N=1-e^{-\frac{1}{3}|\alpha_1-\alpha_2|^2}$ and equals that of the optimal beam-splitter setup from Paper I. However, in contrast to the setup from Paper I, if a conclusive result is obtained before measuring the $N$-th triple we still have ``diluted'' input states at our disposal.

\subsection{More types of reference states}
In the previous section the optimal values transmittivities in our beam-splitter setup were state-independent only in the case of equal number of copies per reference state. Thus,  for more than two types of reference states we will discuss only cases with the same number of copies of each reference state. Unfortunately, we will see that even in this restricted scenario, the optimal choice of transmittivities in the setup we propose will depend on the reference states.

The generalization of the beam-splitter unambiguous identification scheme from the previous subsection is straightforward. We start by preparing  the ``concentrated'' states $\ket{\sqrt{\na}\nstav},\ket{\sqrt{\nb}\alpha_1},\ldots, \ket{\sqrt{\nb}\alpha_M}$. We use $M-1$ beam-splitters to sequentially split the ``concentrated'' unknown state $\ket{\sqrt{\na}\nstav}$ into $M$ states. Each of these $M$ states is then merged with one of the ``concentrated'' reference states $\ket{\sqrt{\nb}\alpha_1},\ldots, \ket{\sqrt{\nb}\alpha_M}$ on beam-splitter $C_1, \ldots, C_M$. The transmittivity $T_k$ (of beam-splitter $C_k$) is chosen so that destructive interference yields vacuum on the second output port of $C_k$ for $\nstav=\alpha_k$. These output ports are monitored by photodetectors $D_1, \ldots, D_M$. Detection of at least one photon by photodetector $D_k$ unambiguously indicates $\nstav \neq \alpha_k$. If all photodetectors except the $k$-th fire, then we conclude that $\nstav=\alpha_k$. For $M=2$ we had freedom in choosing the ratio $T_1$ with which the ``concentrated'' unknown state $\ket{\sqrt{\na}\nstav}$ was split into two parts used for the two comparisons. In order to maximize the probability of identification we can tune $M-1$ transmittivities of the beam-splitters that result in splitting the ``concentrated'' unknown state. The optimal choice of these transmittivities even for equal prior probabilities $\eta_j=1/M$ depends on the choice of the reference states. Once we consider $\nb=\nc=\ldots$ then let us consider equal splitting of the ``concentrated'' unknown state into $M$ parts, even though it is not necessarily the optimal choice. In such a case the beam-splitters $C_1, \ldots, C_M$ are performing the following transformation:
\begin{eqnarray}
C_k &:& \ket{\sqrt{\frac{\na}{M}}\nstav}\otimes\ket{\sqrt{\nb}\alpha_k}\mapsto \ket{out1}\otimes\ket{out2}; \nonumber\\
&\ &\ket{out1}=\ket{\sqrt{\frac{T_k \na}{M}}\nstav+\sqrt{R_k \nb}\alpha_k}; \\
&\ &\ket{out2}=\ket{-\sqrt{\frac{R_k \na}{M}}\nstav+\sqrt{T_k \nb}\alpha_k}.\nonumber
\end{eqnarray}
The condition of $\ket{out2}$ being a vacuum for $\nstav=\alpha_k$ forces us to set the transmittivity to $T_k=\na/(\na+M\nb)$. The probability of observing at least one photon in $\ket{out2}$ if $\nstav=\alpha_j$ is $1-e^{-\frac{ \na\nb}{\na+M\nb}|\alpha_j-\alpha_k|^2}$. The corresponding probability of identification therefore reads:
\begin{eqnarray}
\puim=\sum^{M}_{j=1}\frac{1}{M} \prod_{k\neq j}(1-e^{-\frac{\na \nb}{\na+M\nb}|\alpha_j-\alpha_k|^2}) \, .\nonumber\\
\end{eqnarray}
Let us note that for a single copy of an unknown state and a single copy of a reference state ($\na=\nb=\nc=\ldots=1$) the scenario is the same as in the Sec.~IV.E of Paper I. In that special case both proposed setups coincide, and therefore also the previous expression reduces to Eq. (4.30) from Paper I.

\section{Recovery of reference states after the measurement}
\label{recov1}
In this section we examine the information that remains in the unmeasured modes of our beam-splitter UI setups. In particular, we focus on the possibility of ``recreating'' the reference states out of those modes. This can be useful for creating a quantum database, which would not be completely destroyed by the search performed on it. Instead, the data i.e. the reference states would degrade gradually with repeated use. First, we show that reference states can not be "recreated" without additional resources if the first unambiguous identification yields an inconclusive outcome. Although, this may seem disappointing, we show that the unmeasured states still can be used efficiently for UI if the same unknown state is expected. Next, we examine the situation of the first UI producing a conclusive result known to us. In that case  ``diluted'' reference states can be created, and they can be used for another independent unambiguous identification.

Let us consider the basic version of unambiguous identification of coherent states ($M=2, \na=\nb=\nc=1$). The beam-splitter setup for this scenario was originally proposed in Paper I and coincides with the setup depicted in Fig.~\ref{multi1}. Modes $B$ and $D$ are not entangled with other modes, therefore their state does not depend on the measurement performed by the two photodetectors. The states of the modes $B, D$ is given by the Eqs. (\ref{t2set}), (\ref{bs23stav}), where $T_1$ is set to $1/2$ (for details see section \ref{tworstates}).
\begin{eqnarray}
\ket{out}_B=\ket{\sqrt{\frac{2}{3}}\alpha_1+\sqrt{\frac{1}{6}}\nstav}_B \nonumber\\
\ket{out}_D=\ket{\sqrt{\frac{1}{6}}\nstav+\sqrt{\frac{2}{3}}\alpha_2}_D \label{stavafter}
\end{eqnarray}
Using beam-splitters, phase shifters and known coherent states we can produce out of states Eq. (\ref{stavafter}) a coherent state of the form
\begin{eqnarray}
\ket{a(\sqrt{\frac{2}{3}}\alpha_1+\sqrt{\frac{1}{6}}\nstav)+b(\sqrt{\frac{1}{6}}\nstav+\sqrt{\frac{2}{3}}\alpha_2)+\gamma},
\label{r11}
\end{eqnarray}
where $a,b,\gamma \in \complex$. Imagine we want to recover the first reference state. Hence, we want the state from Eq. (\ref{r11}) to be $\ket{\lambda \alpha_1}$. Even though we know that either $\nstav=\alpha_1$, or $\nstav=\alpha_2$, a suitable choice of $a,b$ for one of these possibilities produces ``junk'' in the other case. Analogous reasoning works for the second reference state. For the inconclusive result of UI measurement we do not know, which possibility took place, and thus the reference states can not be recovered.

\subsection{Repetition of UI for same unknown state}
\begin{figure}
\begin{center}
\includegraphics[width=7cm]{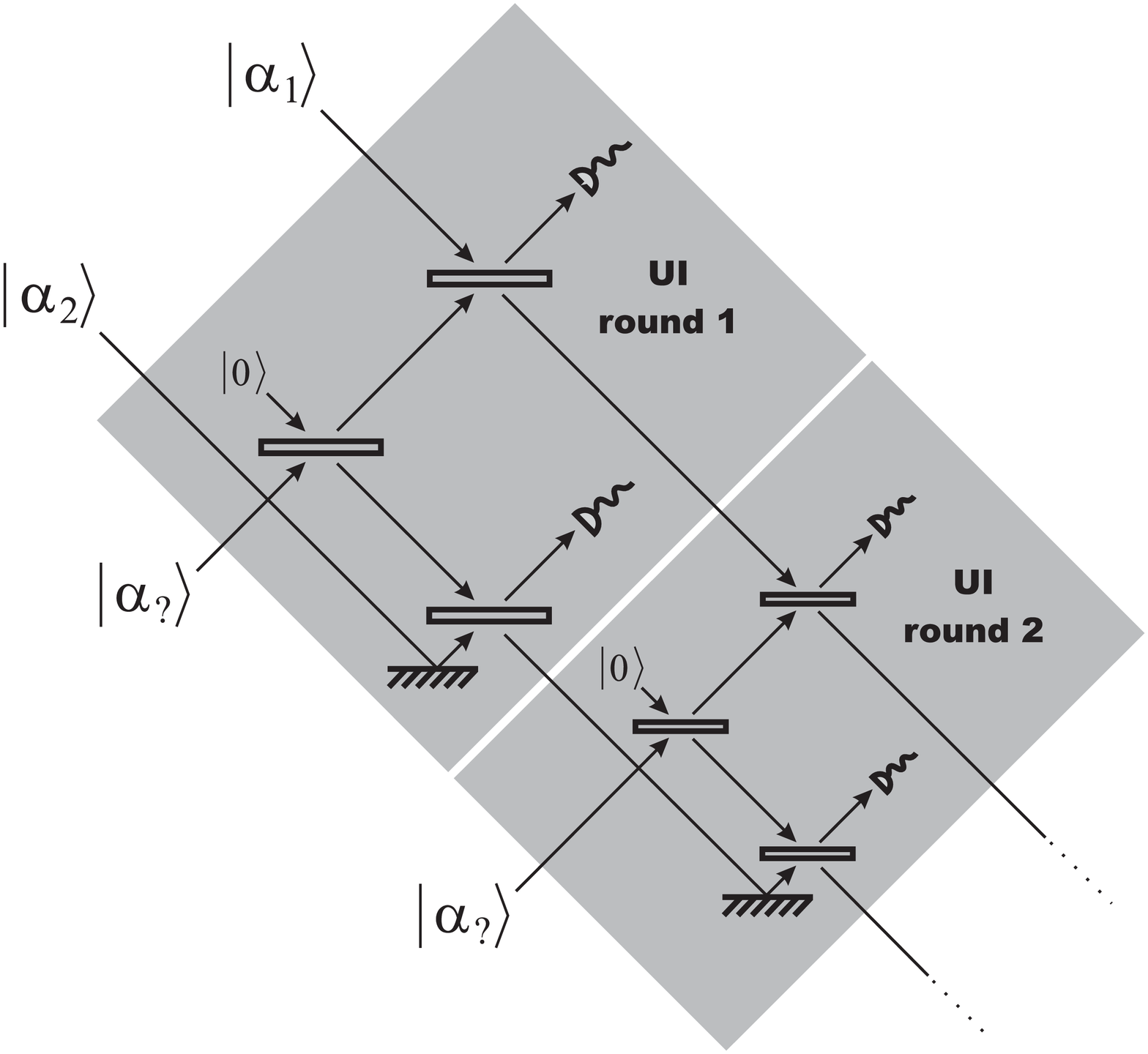}
\caption{The beam-splitter setup designed for a subsequent unambiguous identification of multiple copies of an unknown coherent state.}
\label{recsetup1}
\end{center}
\end{figure}

Although, the unmeasured modes of the beam-splitter setup seem useless they can be exploited in the UI of the same unknown state $\ket{\nstav}$. Namely, we can feed them instead of reference states into the beam-splitter scheme shown in Fig.~\ref{multi1}. The concatenation is illustrated in Fig.~\ref{recsetup1}.
The transmittivity of the beamsplitter $B_2$ (respectively, $B_3$) can be set so that its measured output is in a vacuum if $\nstav=\alpha_1$ (respectively, if $\nstav=\alpha_2$). If we chose (for symmetry reasons) $T_1=1/2$ then the transmittivities $T_2$, $T_3$ should be set to $T_2=3/4$, $T_3=1/4$. This implies that the photodetectors measure the states $\ket{(\nstav-\alpha_1)/\sqrt{6}}$, $\ket{(\alpha_2-\nstav)/\sqrt{6}}$. Thus, for both cases $\nstav=\alpha_1, \nstav=\alpha_2$ we can observe a photon in only one of the photodetectors and with the probability $1-e^{-\frac{1}{6}|\alpha_1-\alpha_2|^2}$ unambiguously conclude which possibility took place. Hence, the probability $1-e^{-\frac{1}{6}|\alpha_1-\alpha_2|^2}$ is a conditional UI probability after a first identification measurement returned an  inconclusive result. The overall probability of an unambiguous identification for this two-round measurement is $1-e^{-\frac{1}{2}|\alpha_1-\alpha_2|^2}$. This is due to the fact that the measurement fails only if both measurement rounds yield an inconclusive outcome.

The two-round measurement is essentially an UI scheme for $M=2$, $\na=2$, $\nb=\nc=1$, so we can compare its performance with the corresponding beam-splitter scheme  Eq. (\ref{puibs}) analyzed in Sec.~\ref{tworstates}. Indeed, the performance is the same, but the two round measurement has one possible advantage. If the first round gives a conclusive result then we still have an unmeasured copy of the unknown state (i.e. copy of $\ket{\alpha_1}$ respectively $\ket{\alpha_2}$) at our disposal.
 This is a similar advantage as in the case of weak implementation of the UI measurement discussed in Sec.~\ref{weak1}.

\subsection{Repetition of UI with different unknown state}
As we illustrated in the beginning of Sec.~\ref{recov1} it is not possible to ``recreate'' the reference states by linear optics after an inconclusive result of an UI measurement is obtained. On the contrary, we will show that when a conclusive result is registered then  both reference states can be ``recreated''. Although, this recreation is not perfect, the recreated reference states are bit ``diluted''. Nevertheless, subsequently, these states can be used as reference states for an UI with a different, independently prepared unknown state $\ket{\beta_?}$ (either $\beta_?=\alpha_1$ or $\beta_?=\alpha_2$).

When the $\nstav=\alpha_1$ result is found in the first round of the UI, the unmeasured modes $B, D$ are in states $\ket{\sqrt{\frac{3}{2}}\alpha_1}_B$ and $\ket{\sqrt{\frac{1}{6}}\alpha_1+\sqrt{\frac{2}{3}}\alpha_2}_D$, respectively. Thus, we have the ``concentrated'' first reference state $\ket{\sqrt{\frac{3}{2}}\alpha_1}$ in the mode B. Let us now examine whether the reference state $\ket{\alpha_2}$ can be ``recreated'' out of the modes $B$ and $D$. The natural idea is to use the mode $B$ to shift the mode $D$ via a beam-splitter so that the $\alpha_1$ part of the amplitude in $\ket{\sqrt{\frac{1}{6}}\alpha_1+\sqrt{\frac{2}{3}}\alpha_2}_D$   is canceled. This happens for the transmittivity of the beam-splitter equal to $9/10$:
\begin{eqnarray}
\ket{\sqrt{\frac{3}{2}}\alpha_1}\otimes\ket{\sqrt{\frac{1}{6}}\alpha_1+\sqrt{\frac{2}{3}}\alpha_2}\mapsto \nonumber\\
\mapsto\ket{\left(\sqrt{\frac{27}{20}}+\sqrt{\frac{1}{15}}\right)\alpha_1+\sqrt{\frac{1}{60}}\alpha_2}\otimes\ket{\sqrt{\frac{3}{5}}\alpha_2} \, .
\end{eqnarray}
\begin{figure}
\begin{center}
\includegraphics[width=6cm]{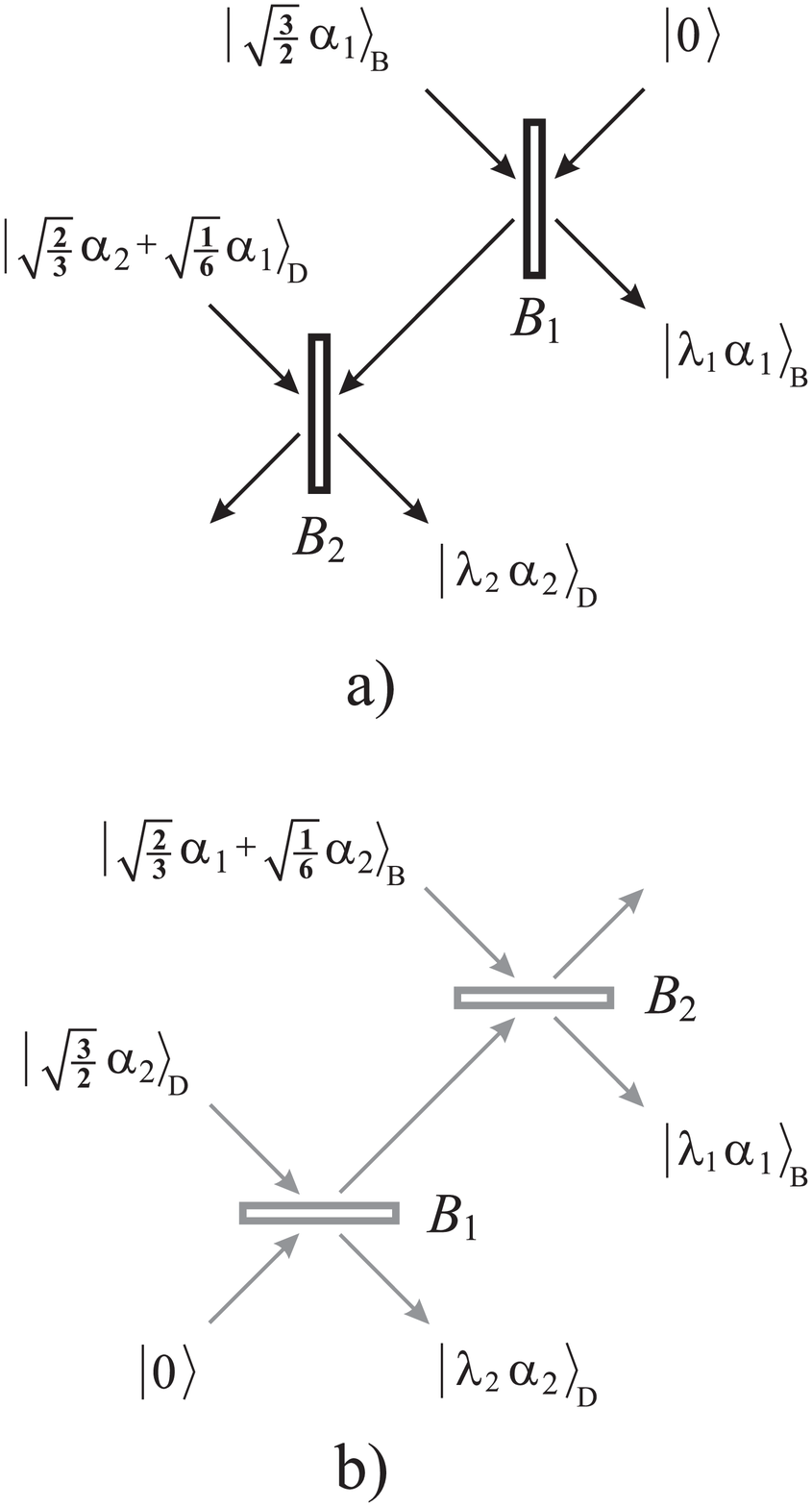}
\caption{The beam-splitter setups designed for the recovery of unmeasured modes from Fig.~\ref{multi1}. In the case $\nstav=\alpha_1$ the setup a) is used, however for $\nstav=\alpha_2$ the setup b) is used.}
\label{recover1}
\end{center}
\end{figure}
Hence, we know how to recover separately either the first or the second reference state. If solely such a single state is used in the subsequent UI measurement then the  probability of success is bounded from above by $1/2$, because only one type of a reference state can be identified. Thus, we want to find a setup, which extracts both types of reference states simultaneously and allows for a subsequent round of the unambiguous identification of $\ket{\beta_?}$. Such a scheme is presented in Fig. \ref{recover1}a. The beam-splitter $B_1$ splits the ``concentrated'' first reference state into two parts. One part can be directly used for the next round of UI, the second part cancels the $\alpha_1$ contribution in the amplitude of the coherent state in the mode $D$ via the beam-splitter $B_2$. If we set transmittivity of the beam-splitter $B_1$ to be $T^R_1$, then the requirement of cancelation of the $\alpha_1$ contribution of the amplitude of the coherent state in the mode $D$ constrains the transmittivity of $B_2$ to be $T^R_2=(9-9 T^R_1)/(10-9 T^R_1)$. The corresponding ``recreated'' reference states then read
\begin{eqnarray}
\ket{\sqrt{\frac{3}{2}T^R_1}\alpha_1}, \quad \ket{\sqrt{\frac{6-6 T^R_1}{10-9 T^R_1}}\alpha_2}\, .
\label{staft1}
\end{eqnarray}
We want to use these two states instead of the reference states $\ket{\alpha_1}$, $\ket{\alpha_2}$ in the next round of UI. Both possible preparations $\ket{\beta_?}=\ket{\alpha_1}$, $\ket{\beta_?}=\ket{\alpha_2}$ will be equally likely, therefore we chose $T^R_1=(7-\sqrt{13})/9$ so that equally diluted reference states
\begin{eqnarray}
\ket{\sqrt{\lambda_2}\alpha_1}, \quad \ket{\sqrt{\lambda_2}\alpha_2}, \quad \lambda_2\equiv\frac{7-\sqrt{13}}{6}
\label{staft2}
\end{eqnarray}
enter the next round of UI. If a conclusive result $\nstav=\alpha_2$ is obtained in the first round of UI, then after exchanging the roles of the modes $B$ and $D$, analogous recovery setup (see Fig.\ref{recover1}b) can be used to produce the ``diluted'' reference states Eq. (\ref{staft2}). Thus, for both conclusive results from the first round of UI, one type of UI measurement using the recovered reference states can be used in the second round.
\begin{figure}
\begin{center}
\includegraphics[width=8.5cm]{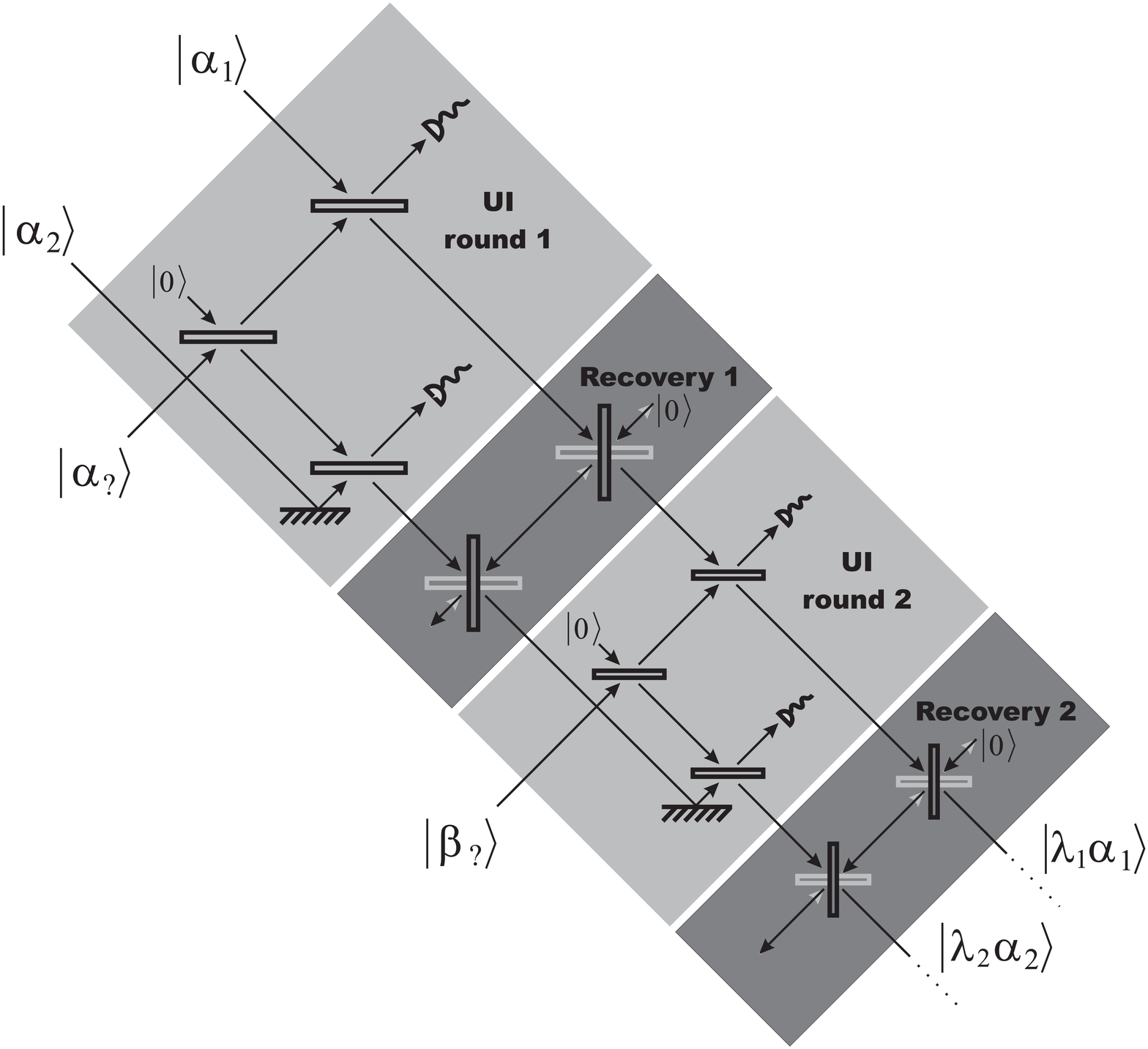}
\caption{The beam-splitter setup designed for repetition of UI with different unknown state, which can be seen as a repeated search in a quantum database. The gray beam-splitters in recovery steps are used if the unknown state from previous round of UI matches the second reference state otherwise black ones are used.}
\label{recsetup2}
\end{center}
\end{figure}
Actually, the beam-splitter setup from Fig. \ref{multi1} can be used (see Fig. \ref{recsetup2}) if we take into account that for our input states $\na=1$, $\nb=\nc=(7-\sqrt{13})/6$. Upon making this substitution the performance of the setup is the same as in Sec.~\ref{tworstates}, and all the formulas derived there remain valid. The aforementioned setup succeeds in UI with the probability given by Eq. (\ref{puibs}). However, the second round of UI will be possible only if the first UI succeeded, which implies the following probability of UI in the second round
\begin{eqnarray}
\pui{}{\alpha_1}{\alpha_2}&=&(1-e^{-\frac{1}{3}|\alpha_1-\alpha_2|^2}) \nonumber\\
&\times&(1-e^{-\frac{7-\sqrt{13}}{2(10-\sqrt{13})}|\alpha_1-\alpha_2|^2}) \, .
\end{eqnarray}
It is interesting that UI with  nearly orthogonal reference states can be done also in the second round with a probability of success  approaching unity.

\begin{figure}
\begin{center}
\includegraphics[width=8.5cm]{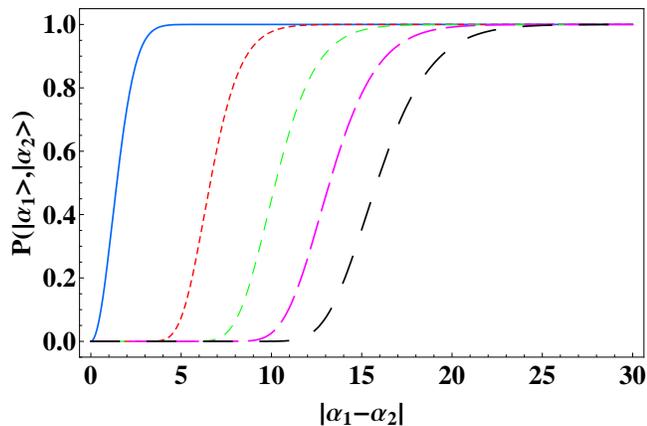}
\caption{(Color online) The performance of the recovery setup.
 The probability
of identification $P(|\alpha_1\rangle,|\alpha_2\rangle)$ as a function of the scalar product
(given by $|\alpha_1-\alpha_2|$) depicted for various numbers of measurement rounds.
Starting from the left the curves correspond to the probability of identification in the first,  20th, 40th, 60th, 80th round of the UI.}
\label{recover2}
\end{center}
\end{figure}

Let us now see, whether further rounds of UI are still possible. The first round of UI can be seen as use of the beam-splitter setup from Fig. \ref{multi1} with $\na=\nb=\nc=1$ followed by the setup from Fig. \ref{recover1} recovering the reference states. In the second round we have used again the beam-splitter setup from Fig. \ref{multi1} this time with $\na=1,\ \nb=\nc=(7-\sqrt{13})/6$. It turns out that we can perform infinitely many additional rounds of UI, where in each round the unknown state is independently chosen to be either $\ket{\alpha_1}$ or $\ket{\alpha_2}$. It suffices to use the beam-splitter setup from Fig. \ref{multi1} followed by the setup from Fig. \ref{recover1} recovering the reference states in each round of UI. However, the transmittivities of the beam-splitters used in those setups must be set as follows. Let us denote by $\sqrt{\lambda_k}$ the factor by which the reference states are suppressed at the beginning of the $k$-th round (e.g. $\lambda_1=1$).  In $k$-th round of UI we should set $T_1=1/2,\  T_2=2 \lambda_k /(1+2 \lambda_k),\  T_3=1/(1+2\lambda_k)$ in the scheme from Fig. \ref{multi1} and
\begin{eqnarray}
T_1&=&1-\frac{2\lambda^2_k+\sqrt{4\lambda^4_k+(1+2\lambda_k)^2}}{(1+2\lambda_k)^2}; \nonumber\\
T_2&=&\frac{(1-T_1)(1+2\lambda_k)^2}{1+(1-T_1)(1+2\lambda_k)^2},
\end{eqnarray}
in the scheme presented in Fig. \ref{recover1}. The suppression of the amplitude of reference states is given by $\lambda_{k}\mapsto \lambda_{k+1}=f(\lambda_k)$, where
\begin{eqnarray}
f(x)=\frac{(1+2x)^2-2x^2-\sqrt{4x^4+(1+2x)^2}}{2(1+2x)}. \nonumber\\
\end{eqnarray}
The probability of successfully performing the UI in the $k$-th round is $P^{(k)}(\ket{\alpha_1},\ket{\alpha_2})=P^{(k-1)}(\ket{\alpha_1},\ket{\alpha_2})(1-e^{-\frac{\lambda_k}{1+2\lambda_k}|\alpha_1-\alpha_2|^2})$, because the $k$-th round of the UI is possible only if all previous UI succeeded \cite{recoverr}. The dependence of the probability of identification on the difference of the amplitudes of the reference states and on the number of measurement rounds is depicted on Fig. \ref{recover2}.

Let us now discuss an alternative approach to the recovery of reference states. Imagine that our task is to identify $N$ independent unknown states with reference states. Instead of recovering reference states after identifying each of the unknown states we can first split the reference states into $N$ parts and then perform the identifications independently. We are going to illustrate that even though we know value $N$ ahead of time, the splitting strategy does not outperform the strategy based on recovery of reference states.

\begin{figure}
\begin{center}
\includegraphics[width=8.5cm]{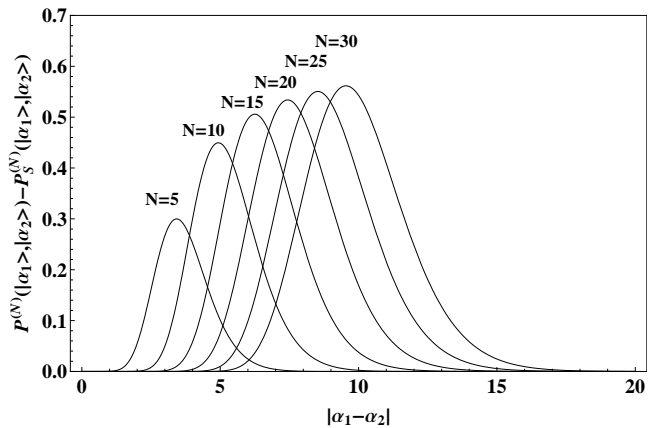}
\caption{The difference between the performance of the recovery and the splitting strategies for different number of identification rounds $N$ as a function of the scalar product (given by $|\alpha_1-\alpha_2|$).}
\label{recover3}
\end{center}
\end{figure}

The splitting strategy begins by distributing the information in the two reference states into $N$ copies of the states $\ket{\frac{1}{\sqrt{N}}\alpha_1}, \ket{\frac{1}{\sqrt{N}}\alpha_2}$. These two states are then put together with one of the unknown states and are unambiguously identified by the scheme for $M=2, \na=1, \nb=\nc=1/N$. The probability of a successful identification the unknown state depends only on the reference states, hence for each of the $N$ UI measurements we have $\pui{}{\alpha_1}{\alpha_2}=1-e^{-\frac{1}{N+2}|\alpha_1-\alpha_2|^2}$. The probability that all of them succeed is therefore
$P^{(N)}_S(\ket{\alpha_1},\ket{\alpha_2})=(1-e^{-\frac{1}{N+2}|\alpha_1-\alpha_2|^2})^{N}$. On the other hand in the scheme with the recovery of the reference states the $N$-th round can succeed only if all the previous identification rounds were successful. This means that the probability of success of the $N$-th round $P^{(N)}(\ket{\alpha_1},\ket{\alpha_2})$ is the same as the probability that all the $N$ rounds of the identification task were successful. The difference between the performance of the recovery and the splitting strategies for different $N$ is depicted in Fig.~\ref{recover3}.

\section{Influence of noise on reliability of UI setups}  
In this section we investigate how noise (uncertainty) in the state preparation affects the reliability of the measurement results. The UI setups we have presented above are designed specifically for coherent states and ideally they are $100\%$ reliable, i.e. whenever we obtain a conclusive result $E_i$ then we can be completely sure that the possibility $x_i$ (i.e., $\nstav=\alpha_i$) took place.However, it might be that the unknown and reference states are sent to us via a noisy channel or simply that their preparation is  noisy. We assume that this disturbance has the form of a technical noise \cite{noise}, and therefore the unknown and the reference states are not pure coherent states $\ket{\alpha_i}$, but rather their mixtures $\omega_i$:
\begin{eqnarray}
\omega_i&=&\frac{1}{2\pi \sigma^2}\int_{\complex}d\beta e^{-\frac{|\beta|^2}{2\sigma^2}} \ket{\alpha_i+\beta} \bra{\alpha_i+\beta}\label{omegai} \,  ;\\
\rho_i(\boldsymbol{\alpha})&=&(\omega_i)^{\otimes\na}\otimes(\omega_1)^{\otimes\nb}\otimes(\omega_2)^{\otimes\nc}\otimes\ldots ,
\end{eqnarray}
with $\sigma$ defining the strength of the noise \cite{integral} and $\boldsymbol{\alpha}$ indicates the dependence on $\alpha_i$.
In such a case conclusive results of our UI setups will no longer be unambiguous. More precisely, there will be a certain probability $P(x_i|E_i)$ with which the obtained outcome $E_i$ of the measurement is the consequence of the possibility $x_i$. This probability is called  the {\it reliability} of the outcome $E_i$. The corresponding mathematical definition reads:
\begin{eqnarray}
R(E_i)=P(x_i|E_i)=\frac{\eta_i P(E_i|x_i)}{\sum_{j=1}^{M}\eta_j P(E_i|x_j)} , \label{rel}
\end{eqnarray}
where $\eta_i$ is the a priori probability of the possibility $x_i$ and $P(E_i|x_j)$ is the probability that the measurement of the system prepared in the possibility $x_j$ will give a result $E_i$. Let us note that under the possibility $x_i$ we understand all situations in which the unknown state is the same as the $i$-th reference state. Thus $x_i$ stands for the whole set of situations, which differ by complex amplitudes $\alpha_k$ of the ``centers'' of the reference states $\omega_k$. 
 How those ``center points'' of all reference states are chosen in $x_i$ is described by the probability distribution $\chi_i(\alpha_1,\ldots,\alpha_M)$. The support of $\chi_i$ is $\complex^m$ corresponding to an infinite plane. Therefore a uniform probability distribution can not be defined on it. Nevertheless, we can express the reliability as:
\begin{eqnarray}
R(E_i)=\frac{\eta_i \int_{\complex^M} d\boldsymbol{\alpha} \chi_i(\boldsymbol{\alpha}) Tr(E_i\rho_i(\boldsymbol{\alpha}))}{\sum_{j=1}^{M}\eta_j  \int_{\complex^M}d\boldsymbol{\alpha} \chi_i(\boldsymbol{\alpha}) Tr(E_i\rho_j(\boldsymbol{\alpha}))}, \label{relstredna}
\end{eqnarray}
where $d\boldsymbol{\alpha}\equiv d\alpha_1\ldots d\alpha_M$. In the limit $\sigma\rightarrow 0$ states $\omega_i$ become $\ket{\alpha_i}\bra{\alpha_i}$. Because of the no-error conditions (\ref{stavy2}),  which for coherent states are satisfied by our UI setups, only the $i$-th term of the sum in Eq. (\ref{rel}) survives. Thus, without noise the reliability is equal to unity. For $\sigma>0$ also other terms in Eq. (\ref{rel}) will contribute and hence reliability will be less than one. Moreover, the precise value of $R(E_i)$ will depend on the probability distributions $\chi_i(\boldsymbol{\alpha})$.

In the remaining part of this section we will investigate a scenario, which might be called as the {\it phase keying}. We assume that two reference states ($M=2$) have always opposite phases, i.e. if $\omega_1$ is centered around the amplitude $\alpha$ then $\omega_2$ is centered around the amplitude $-\alpha$. Values of $\alpha$ have a Gaussian distribution centered around $0$ (vacuum) with a dispersion $\xi$ , so
\begin{eqnarray}
\chi_i(\alpha_1,\alpha_2)=\delta(\alpha_1+\alpha_2)\frac{1}{2\pi \xi^2}e^{-|\alpha_1|^2/(2\xi^2)},\quad i=1,2.\nonumber\\
\label{chidistr}
\end{eqnarray}

In order to calculate the reliability we must first evaluate $Tr[E_i\rho_j(\boldsymbol{\alpha})]$. This means we have to derive the probabilities with which detectors $D_1, D_2$ click if ``fuzzy'' states $\omega_?, \omega_1, \omega_2$ are fed into the UI setup instead of $\ket{\nstav}, \ket{\alpha_1}, \ket{\alpha_2}$.
Our UI setup uses an additional mode $D$ that should be initially prepared in vacuum. We assume that also this mode is noisy and initially in a state $\omega_i$ centered around $0$ (vacuum).

To present our calculations concisely, we first derive how the setup acts on coherent states: We integrate over coherent states 
in Eq. (\ref{omegai}) (e.g. $\ket{\alpha_i+\beta}$) and then we integrate those partial results. Thus, for a single copy of the unknown and the reference states we derive how the UI setup acts on states $\ket{\nstav+\nu}, \ket{\alpha_1+\beta}, \ket{\alpha_2+\gamma},\ket{\varrho}$ fed into modes $A,B,C,D$ (see Fig.~\ref{genstrat1}) and finally we perform  integration over $\nu, \beta, \gamma, \varrho$.

For multiple copies of the unknown and the reference states we assume that the noise is acting independently on each of the copies, i.e. we would analyze $\nb$ copies of the first reference state entering as states $\ket{\alpha_1+\beta_1},\ldots,\ket{\alpha_1+\beta_{\nb}}$. The first part of the UI setup, which ``concentrates'' copies of the same species, generates the state $\ket{\sqrt{\nb}\alpha_1+\frac{1}{\sqrt{\nb}}(\beta_1+\ldots+\beta_{\nb})}$ and similarly, the state $\ket{\sqrt{\nc}\alpha_2+\frac{1}{\sqrt{\nc}}(\gamma_1+\ldots+\gamma_{\nc})}$ for the second reference state, and $\ket{\sqrt{\na}\nstav+\frac{1}{\sqrt{\na}}(\nu_1+\ldots+\nu_{\na})}$ for the unknown state. The beam splitter transformation for coherent input states does not entangle its outputs, thus we can, in the same way as in section \ref{tworstates}, derive expressions for the states of the modes that the photodetectors $D_1, D_2$ measure. Consequently, the final states of the modes $A$ and $C$ read:
\begin{eqnarray}
\Big|\sqrt{\frac{\na\nb}{\na+2\nb}}\left[\nstav-\alpha_1-\sqrt{\frac{1}{\na}}\varrho+\frac{1}{\na}\boldsymbol{\nu}-\frac{1}{\nb}\boldsymbol{\beta} \right]\Big>_A\nonumber\\
\equiv \ket{\mu_1}_A \; \nonumber\\
\\
\Big|\sqrt{\frac{\na\nc}{\na+2\nc}}\left[\alpha_2-\nstav-\sqrt{\frac{1}{\na}}\varrho-\frac{1}{\na}\boldsymbol{\nu}+\frac{1}{\nc}\boldsymbol{\gamma} \right]\Big>_C \nonumber\\
\equiv \ket{\mu_2}_C , \nonumber
\end{eqnarray}
where $\boldsymbol{\nu}\equiv\sum_{k=1}^{\na}\nu_k$, $\boldsymbol{\beta}\equiv\sum_{k=1}^{\nb}\beta_k$, $\boldsymbol{\gamma}\equiv\sum_{k=1}^{\nc}\gamma_k$.
Now we have to evaluate the probability of the projection of these states $\ket{\mu_1}_A,\ket{\mu_2}_C$ onto the vacuum. Subsequently, we will integrate this partial result to obtain the probability $P(D_k|\rho_i(\boldsymbol{\alpha}))$ that the photodetector $D_k$ ($k=1,2$) does not click. Probabilities $P(D_k|\rho_i(\boldsymbol{\alpha}))$ are related to $Tr(E_i\rho_j(\boldsymbol{\alpha}))$ in the following way:
\begin{eqnarray}
Tr(E_1\rho_1)&=&[1-P(D_1|\rho_1)].P(D_2|\rho_1)\, ;\nonumber\\
Tr(E_1\rho_2)&=&[1-P(D_1|\rho_2)].P(D_2|\rho_2)\, ;\nonumber\\
Tr(E_2\rho_1)&=&P(D_1|\rho_1).[1-P(D_2|\rho_1)]\, ;\nonumber\\
Tr(E_2\rho_2)&=&P(D_1|\rho_2).[1-P(D_2|\rho_2)]\,  \label{treiroj},
\end{eqnarray}
where the argument of $\rho_i(\boldsymbol{\alpha})$ is omitted for brevity.
Finally, we obtain the quantities $Tr[E_i\rho_j(\boldsymbol{\alpha})]$ that we need for evaluating the reliability according to Eq. (\ref{relstredna}).

\begin{figure}
\begin{center}
\includegraphics[width=8.5cm]{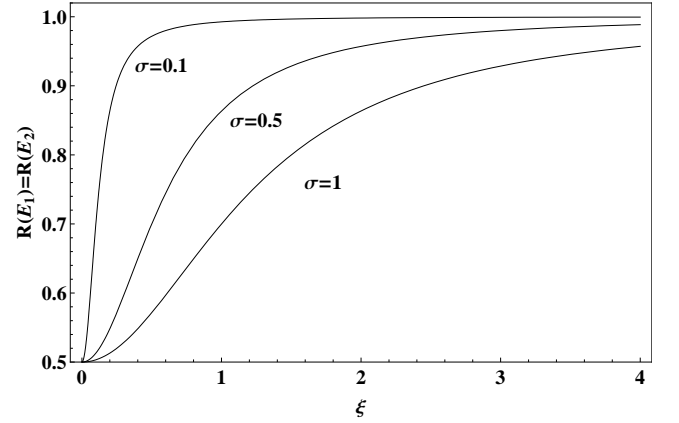}
\caption{The reliability of the UI setup ($M=2$, $\na=\nb=\nc=1$) as a function of the typical displacement $\xi$.
Different curves correspond to different values of $\sigma$ i.e. to different fuzziness of the states. As is seen from the figure all curves
in the limit of large $\xi$ are approaching the unity.}
\label{relgraf1}
\end{center}
\end{figure}

\begin{figure}
\begin{center}
\includegraphics[width=8.5cm]{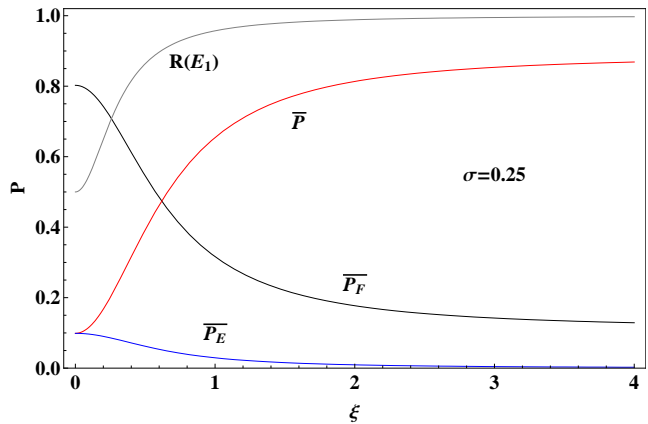}
\caption{The reliability and the average probability of success ($\mpui{}$), the error ($\mpui{E}$), and the failure ($\mpui{F}$) for the ``phase keying'' scenario ($M=2$, $\na=\nb=\nc=1$) with $\sigma=0.25$ as a function of the typical displacement $\xi$ .}
\label{rpppgraf}
\end{center}
\end{figure}

Using the formula $|\Bra{0}\ket{\mu_i}|^2=e^{-|\mu_i|^2}$ for the modulus of the overlap of two coherent states we obtain:
\begin{widetext}
\begin{eqnarray}
P(D_1|\rho_i(\boldsymbol{\alpha}))=
\int_{\complex^{m}}\frac{d\varrho d\boldsymbol{\gamma}d\boldsymbol{\nu}}{(2\pi\sigma^2)^{m}}\ exp\Big[-\frac{|\varrho|^2+\sum_{k=1}^{\na}|\nu_k|^2+\sum_{k=1}^{\nb}|\gamma_k|^2}{2\sigma^2}
-\frac{\na\nc  \Big| \alpha_2-\nstav-\sqrt{\frac{1}{\na}}\varrho-\frac{1}{\na}\boldsymbol{\nu}+\frac{1}{\nc}\boldsymbol{\gamma}  \Big|^2 }{\na+2\nc}
\Big];\nonumber\label{int1}\\
\\
P(D_2|\rho_i(\boldsymbol{\alpha}))=
\int_{\complex^{n}}\frac{d\varrho d\boldsymbol{\beta}d\boldsymbol{\nu}}{(2\pi\sigma^2)^{n}}\ exp\Big[-\frac{|\varrho|^2+\sum_{k=1}^{\na}|\nu_k|^2+\sum_{k=1}^{\nb}|\gamma_k|^2}{2\sigma^2}
-\frac{\na\nb  \Big|\nstav-\alpha_1-\sqrt{\frac{1}{\na}}\varrho+\frac{1}{\na}\boldsymbol{\nu}-\frac{1}{\nb}\boldsymbol{\beta} \Big|^2 }{\na+2\nb}
\Big],\nonumber\label{int2}\\
\end{eqnarray}
\end{widetext}
where $m=\na+\nc+1$, $n=\na+\nb+1$. The integrals in Eq. (\ref{int1}) and (\ref{int2}) can be performed using the relations derived in Appendix \ref{appintegral}. The results of the integration read:
\begin{eqnarray}
P(D_1|\rho_i(\boldsymbol{\alpha}))=\frac{1}{1+2\sigma^2}e^{-\frac{1}{1+2\sigma^2}\frac{\na\nc}{\na+2\nc}|\alpha_i-\alpha_2|^2}\, ;\\
P(D_2|\rho_i(\boldsymbol{\alpha}))=\frac{1}{1+2\sigma^2}e^{-\frac{1}{1+2\sigma^2}\frac{\na\nb}{\na+2\nb}|\alpha_i-\alpha_1|^2}\, ,
\end{eqnarray}
where we have used the formulas for the case $x_i$, i.e. $\nstav=\alpha_i$.
Consequently, using these results in Eq. (\ref{treiroj}) we obtain:
\begin{eqnarray}
Tr(E_1\rho_1)&=&\frac{1+2\sigma^2-e^{-\frac{1}{1+2\sigma^2}\frac{\na\nc}{\na+2\nc}|\alpha_1-\alpha_2|^2}}{(1+2\sigma^2)^2};\nonumber\\
Tr(E_1\rho_2)&=&\frac{2\sigma^2}{(1+2\sigma^2)^2}\ e^{-\frac{1}{1+2\sigma^2}\frac{\na\nb}{\na+2\nb}|\alpha_1-\alpha_2|^2};\nonumber\\
Tr(E_2\rho_1)&=&\frac{2\sigma^2}{(1+2\sigma^2)^2}\ e^{-\frac{1}{1+2\sigma^2}\frac{\na\nc}{\na+2\nc}|\alpha_1-\alpha_2|^2};\nonumber\\
Tr(E_2\rho_2)&=&\frac{1+2\sigma^2-e^{-\frac{1}{1+2\sigma^2}\frac{\na\nb}{\na+2\nb}|\alpha_1-\alpha_2|^2}}{(1+2\sigma^2)^2}.\nonumber\\
\label{trero}
\end{eqnarray}
Now in order to obtain the reliability it remains to substitute Eqs. (\ref{chidistr}), (\ref{trero}) into Eq. (\ref{relstredna}) and to perform the remaining integrals. Those integrals can be performed in polar coordinates, where the angular dependence is trivial and the radial part  can be simplified with the help of a substitution $t=e^{-r^2/2}$. After performing the integration we obtain the final result, which can be, for $\nb=\nc$,  written in the compact form:
\begin{eqnarray}
R(E_1)=R(E_2)=\frac{1+\theta}{1+2\theta};\nonumber\\
\theta=\frac{\na+2\nb}{\na\nb}\left(\frac{\sigma}{2\xi}\right)^2 .
\label{relvysled}
\end{eqnarray}
Let us note that $\lim_{\sigma\rightarrow 0}R(E_i)=1$ is as it should. Moreover, the reliability depends only on the fuzziness of the states entering the UI setup $\sigma$, the typical difference of the amplitudes of the reference states $2\xi$ and the number of copies that are available.
If $\sigma\ll\xi$, i.e. the fuzziness of the states, is much smaller than the displacement used to encode the information, then $\theta\rightarrow 0$ and $R(E_i)$ approaches the unity. More quantitative insight in the case of a single copy of the unknown and the reference states is provided by Fig. \ref{relgraf1}.
In order to see how the noise influences other relevant quantities we will calculate $\mpui{}, \mpui{E}, \mpui{F}$, which are called the averaged probability of success, the error, and the failure, respectively. Obviously, we either guess correctly, or incorrectly, or do not guest at all (inconclusive result/failure), therefore $\mpui{}+\mpui{E}+\mpui{F}=1$ must hold. It is useful to rewrite the definition of these quantities in the following form:
\begin{eqnarray}
\mpui{}&=&\frac{1}{2}\sum_{i=1}^{2}\int_{\complex^2}d\boldsymbol{\alpha}Tr(E_i\rho_i(\boldsymbol{\alpha}))\chi_i(\boldsymbol{\alpha})\, ;\nonumber\\
\mpui{E}&=&\frac{1}{2}\int_{\complex^2}d\boldsymbol{\alpha}(Tr(E_2\rho_1(\boldsymbol{\alpha}))+Tr(E_1\rho_2(\boldsymbol{\alpha})) )\chi_1(\boldsymbol{\alpha})\, ; \nonumber\\
\mpui{F}&=&1-\mpui{}-\mpui{E} \, .
\end{eqnarray}
Now it suffice to substitute Eqs. (\ref{trero}) into the above equations and to perform the integration in polar coordinates in the same way as in the previous paragraph. The resulting expressions read:
\begin{eqnarray}
\mpui{}&=&\frac{1}{1+2\sigma^2}(1-\frac{1}{1+2\sigma^2+\frac{8\na\nb}{\na+2\nb}\xi^2})\ ; \nonumber\\
\mpui{E}&=&\frac{1}{1+2\sigma^2}(\frac{2\sigma^2}{1+2\sigma^2+\frac{8\na\nb}{\na+2\nb}\xi^2})\, ;\\
\mpui{F}&=&\frac{2\sigma^2}{1+2\sigma^2}+\frac{1-2\sigma^2}{1+2\sigma^2}\frac{1}{1+2\sigma^2+\frac{8\na\nb}{\na+2\nb}\xi^2}\, . \nonumber\\
\end{eqnarray}
More quantitative insight is presented in Fig.~\ref{rpppgraf}, which for the fixed $\sigma=0.25$ presents the behavior of the calculated quantities $\mpui{}, \mpui{E}, \mpui{F}$ as a function of the typical displacement $\xi$. It is worth mentioning that for $\xi\rightarrow\infty$ the average probability of error goes to zero, but $\mpui{F}>0$, because the noise causes inconclusive results by firing both detectors simultaneously.

\section{Conclusion}
In this paper we have studied a specific discrimination task called the unambiguous identification (UI) of coherent states. In this problem we are given a set of identical quantum systems (modes of electromagnetic field) prepared in coherent states. Some of these coherent states are unknown and some of which serve as reference states. The promise is that one of the reference states is the same as the unknown state and the task is to find out unambiguously which one it is. In Sec.~II we presented a generalization of the optical setup we proposed in Paper I \cite{ui1} to situations with more copies of the unknown and the reference states. Our approach was based on an idea of the ``concentration'' of the same type of states into strong coherent states that were subsequently identified by setups for the single-copy scenario. In the UI task it is assumed that the particular choice of the reference states is unknown to us, and only the probability distribution $\chi$ describing this choice is known. Nevertheless, even without having $\chi$ it is possible to derive the optimal choice of transmittivities in the beam-splitter setup we proposed for two types of reference states and an equal number of copies of each of the reference states ($\nb=\nc$). In that case the probability of identification for the reference states $\ket{\alpha_1}, \ket{\alpha_2}$ reads:
\begin{eqnarray}
\pui{}{\alpha_1}{\alpha_2}=1-e^{-\frac{\na \nb}{\na+2\nb}|\alpha_1-\alpha_2|^2}\, .
\end{eqnarray}
In the limit of $\nb=\nc\rightarrow\infty$ the two reference states become known. Therefore, one needs to unambiguously discriminate the unknown state between two known pure states. The probability of success of our setup in this case coincides with the optimal value achieved by the Ivanovic-Dieks-Peres measurement \cite{ivanovic,dieks,peres}.

In Sec.~III we addressed the question whether the reference states can be recreated after our UI measurement. We showed that the reference states can be partially recovered only if the measurement yielded a conclusive outcome. The recovered reference states can be used in the next round of the UI if another unknown state is provided. This might be seen as a repeated search in a quantum database, where the data, i.e. the reference states, degrade with repeated use of the database.

Recently, a framework for transformations induced by linear optics on coherent states was proposed by B. He, J. Bergou in \cite{bergou}. They illustrated their method on the setup proposed in Paper I and suggested that the reference states can be always perfectly recovered. However, in their case the reference states are known, whereas in our case the complex amplitudes of all coherent states are not known in advance.

In Sec.~IV we investigated how a particular type of noise influenced the reliability of the conclusions drawn by our UI setup. More precisely, we considered a communication scenario called the phase keying, with two coherent reference states of equal amplitude, but the opposite phases. We saw that the reliability of results, expressed by Eq. (\ref{relvysled}), depends only on the ratio of the amplitudes of the noise and the signal. However, for nonzero noise the unambiguity of the conclusions is lost.



\section*{Acknowledgment}
The authors want to thank J.Bergou and B.G.Englert for helpful and stimulating discussions. This work was supported by the European Union projects HIP and QAP, by Slovak grant agencies APVV and VEGA via projects RPEU-0014-06 and 2/0092/09, respectively.

\vskip 1cm
\appendix
\section{Optimality proof}
\label{app1}
In this Appendix we shall prove optimality of the UI setup if only
linear optical elements, number resolving detectors and
sources of multimode coherent states are allowed to be used.
Due to the fact that the linear optical transformations preserve the tensor
product structure of coherent states it follows that in any measurement
(using arbitrarily many photodetectors) the measured state
is a factorized coherent state of $N$ modes of the form
$\ket{\beta_1\otimes\cdots\otimes\beta_N}=\ket{\beta_1}
\otimes\cdots\otimes\ket{\beta_N}\equiv\ket{\vec{\beta}}$.
In order to use an outcome of the measurement for the unambiguous conclusion
the probabilities for all the other options must vanish. Let us notice that
for the considered family of states each photodetector measuring the individual
mode has a nonvanishing probability to observe $n>0$ photons unless
this mode is in the vacuum state, i.e. if $\ket{\beta_j}\neq\ket{0}$, then
$p_n(\ket{\beta_j})=|\langle n|\beta_j\rangle|^2>0$ for all $n>0$. Only for the
vacuum state $p_n(\ket{0})=0$. Moreover, the probability to observe no photon
is nonvanishing for all coherent states, i.e. this event cannot be used
for unambiguous conclusion. Consequently, the unambiguous
conclusions are necessarily associated with observation of the nonzero number
of photons identifying
the fact that the corresponding mode is not in the vacuum state.

In the case of unambiguous identification our goal is to discriminate
two families of states: either $\ket{\alpha_1\otimes\alpha_1\otimes\alpha_2}$,
or $\ket{\alpha_2\otimes\alpha_1\otimes\alpha_2}$, where $\ket{\alpha_1},
\ket{\alpha_2}$ are arbitrary coherent states, but $\alpha_1\neq \alpha_2$.
In general, our (Gedanken) experiment starts with a preparation of a coherent state
$\ket{\alpha_?\otimes\alpha_1\otimes\alpha_2\otimes\beta_1\otimes\cdots}$,
where $\ket{\beta_j}$ are fixed states of some ancillary modes.
By linear optical elements this state is mapped into a state
$\ket{\Delta_0\otimes\Delta_1\otimes\Delta_2\otimes\Delta_3\otimes\cdots}$,
where $\Delta_j$ are complex numbers depending on $\alpha_?,\alpha_1,\alpha_2$.
Each of these modes is measured by a photodetector. In order to make
an unambiguous conclusion $\alpha_?=\alpha_1$ based also on a click of the $j$th photodetector
we need to guarantee for all values of $\alpha_1,\alpha_2$ that
$\Delta_j=0$ for $\alpha_?=\alpha_2$ and
$|\Delta_j|>0$ for $\alpha_?=\alpha_1$. 
Similarly, for the unambiguous conclusion $\alpha_?=\alpha_2$.
As it was shown by He and Bergou \cite{bergou} the linear optical
transformations of coherent states can be described by unitary matrices
acting on vectors of amplitudes of individual modes, i.e.
\begin{eqnarray}
\left( \begin{array}{cccccc}
c_{11} & c_{12} & c_{13} & \ldots & \\
c_{21} & c_{22} & c_{23} & \ldots & \\
& & & & \\
\vdots & \vdots & \vdots & \ddots  & \\
& & & &
\end{array} \right)
\left( \begin{array}{c}
\nstav \\ \alpha_1 \\ \alpha_2 \\ \beta_1 \\ \vdots  \\
\end{array} \right)=
\left( \begin{array}{c}
\Delta_1\\
\Delta_2\\
\\  \vdots  \\
\end{array} \right),
\label{transf1}
\end{eqnarray}
with
\begin{eqnarray}
\ket{\Delta_j}=\ket{c_{j1}\nstav+ c_{j2}\alpha_1+c_{j3}\alpha_2+\gamma_j}
\end{eqnarray}
and $\gamma_j=\sum_k c_{j,k+3}\beta_k$. The condition
$\Delta_j=0$ holding for all values $\alpha_1,\alpha_2$
if $\alpha_?=\alpha_2$ implies $c_{j2}=\gamma_j=0$ and
$c_{j1}=-c_{j3}=\lambda_j$, i.e. $\ket{\Delta^{(1)}_j}=
\ket{\lambda_j(\alpha_?-\alpha_2)}$, where the upper index
indicates the association of observation of photons in this mode with the conclusion $\alpha_?=\alpha_1$.
Similarly, if the $j$th mode
will be associated with the conclusion $\alpha_?=\alpha_2$, then
the corresponding state has to be $\ket{\Delta^{(2)}_j}=
\ket{\lambda_j(\alpha_?-\alpha_1)}$.

The detectors can be divided into three
classes according to the type of states that are measured: i) $\ket{\Delta^{(1)}_j}$ (detecting $\alpha_?=\alpha_1$),
ii) $\ket{\Delta^{(2)}_j}$ (detecting $\alpha_?=\alpha_2$),
and, iii) different type of a state corresponding to an inconclusive result. The detectors from the third class can not be employed in making unambiguous decision and hence will not be considered further. An arbitrary click on the detector i) tells us that $\nstav=\alpha_1$ therefore we associate these clicks with the unambiguous result $\nstav=\alpha_1$. Analogously, clicks from the type ii) detector are associated with the unambiguous result $\nstav=\alpha_2$.
In what follows we shall show that
the events on detectors leading to the same conclusion can be replaced
by a single detector while the success probability is preserved.
In other words, an experiment in which $n_1$ detectors are used to conclude that
$\alpha_?=\alpha_1$ and $n_2$ detectors to detect that $\alpha_?=\alpha_2$
can be replaced by an experiment with only two photodetectors. In particular, by renaming the output ports
the output vector can be rearranged into the form
\begin{equation}
\left(
\begin{array}{c}
\Delta_1^{(1)}\\
\vdots\\
\Delta_{n_1+1}^{(2)}\\
\vdots\\
\Delta_{n_1+n_2+1}\\
\vdots\\
\end{array}
\right)=
\left(\begin{array}{c}
(\alpha_?-\alpha_2)\lambda_1\\
\vdots\\
(\alpha_?-\alpha_1)\lambda_{n_1+1}\\
\vdots\\
\Delta_{n_1+n_2+1}\\
\vdots
\end{array}
\right)\equiv\vec{\alpha}^\prime_? \, .
\end{equation}
In such case we denote $\Omega\equiv e^{-|\alpha_1-\alpha_2|^2}$ and the success probability reads
\begin{eqnarray}
P_{\rm success}&=&\frac{1}{2}(1-\prod_{j=1}^{n_1} e^{-|\lambda_j(\alpha_1-\alpha_2)|^2})\nonumber\\
& &+\frac{1}{2}(1-\prod_{j=n_1+1}^{n_1+n_2} e^{-|\lambda_j(\alpha_1-\alpha_2)|^2})\label{prob1}\\
&=&1-\frac{1}{2}(\Omega^{\sum_{j=1}^{n_1}|\lambda_j|^2}+\Omega^{\sum_{j=n_1+1}^{n_1+n_2}|\lambda_j|^2})\, ,
\nonumber
\end{eqnarray}
because the UI measurement fails only if none of the conclusive detectors fire.
However, there exist a unitary matrix of the block diagonal form
\begin{equation}
U=\left( \begin{array}{ccc}
U_1& O  & O \\
O  & U_2& O \\
O  & O  & I
\end{array}\right)\, ,
\end{equation}
where $U_1,U_2$ are suitable unitary matrices $n_i\times n_i$
such that
\begin{eqnarray}
U_1&:&(\lambda_1,\ldots,\lambda_{n_1})^T \mapsto (\kappa_1,0,\dots,0)^T \, ;\nonumber\\
U_2&:&(\lambda_{n_1+1},\ldots,\lambda_{n_1+n_2})^T\mapsto (\kappa_2,0,\dots,0)^T \, .\nonumber
\end{eqnarray}
with $\kappa_1=\sqrt{\sum_{k=1}^{n_1} |\lambda_k|^2}$
and $\kappa_2=\sqrt{\sum_{k=1}^{n_2}|\lambda_{n_1+k}|^2}$.
This means that the overall product of coherent states transforms into
\begin{eqnarray}
U: \vec{\alpha}_?^\prime&\mapsto& \left(\begin{array}{c}
\kappa_1(\alpha_?-\alpha_2)\\
0\\
\vdots\\
\kappa_2(\alpha_?-\alpha_1)\\
0\\
\vdots\\
\Delta_{n_1+n_2+1}\\
\vdots
\end{array}
\right)\, .
\end{eqnarray}
Two detectors measuring the first and the $(n_1+1)^{\rm th}$ output port are of the first respectively the second type and we see that the probability of success
\begin{eqnarray}
P_{\rm success}&=&\frac{1}{2}(1-e^{-|\kappa_1(\alpha_1-\alpha_2)|^2})+\frac{1}{2}(1-e^{-|\kappa_2(\alpha_2-\alpha_1)|^2})\nonumber\\
&=&1-\frac{1}{2}(\Omega^{\sum_{j=1}^{n_1}|\lambda_j|^2}+\Omega^{\sum_{j=n_1+1}^{n_1+n_2}|\lambda_j|^2})
\end{eqnarray}
equals the multidetector case (see Eq.[\ref{prob1}]). This means we have shown that it suffice to consider one conclusive photodetector of the type one and one of the type two. We can now go back to Eq. (\ref{transf1}) and require that
the states measured by the photodetectors $D_1, D_2$ have the form $\ket{\Delta_1}=\ket{\lambda_1(\nstav-\alpha_2)}, \ket{\Delta_2}=\ket{\lambda_2 (\nstav-\alpha_1)}$. This corresponds to the following transformation matrix from Eq. (\ref{transf1})
\begin{eqnarray}
W=\left( \begin{array}{cccc}
\lambda_1 & 0 & -\lambda_1 & \ldots \\
\lambda_2 & -\lambda_2 & 0 & \ldots \\
\vdots & \vdots & \ddots  &
\end{array} \right).
\label{transf2}
\end{eqnarray}
Let us now find the bounds on $|\lambda_1|, |\lambda_2|$ required by
the unitarity of the matrix $W$. At first, each row is normalized, i.e.
\begin{eqnarray}
1&=&\sum_i |c_{1i}|^2=2|\lambda_1|^2+a^2=\sum_i |c_{2i}|^2=2|\lambda_2|^2+b^2\, ,\nonumber
\end{eqnarray}
where $a,b$ are norms of remaining parts of the first and the second row vectors, respectively. Their orthogonality and the Cauchy-Schwartz inequality give us the inequality $|\lambda_1\lambda_2|\leq ab$. With the help of the previous equation we find
\begin{eqnarray}
|\lambda_1|^2|\lambda_2|^2\leq (1-2|\lambda_1|^2)(1-2|\lambda_2|^2).
\label{inq1}
\end{eqnarray}
The probability of success in the UI for the scheme using linear optical
elements described by the matrix $W$ is
\begin{eqnarray}
\pui{}{\alpha_1}{\alpha_2}=\frac{1}{2}\sum_{i=1}^2 (1-e^{-|\lambda_i|^2 |\alpha_1-\alpha_2|^2}).
\end{eqnarray}
The higher the $|\lambda_i|$'s the higher $\pui{}{\alpha_1}{\alpha_2}$ is. However, the values of $\lambda_1,\lambda_2$ must satisfy the inequality
(\ref{inq1}) and therefore the maximum is limited to
\begin{eqnarray}
\pui{}{\alpha_1}{\alpha_2}&=&\frac{1}{2}(1-e^{-|\lambda_1|^2|\alpha_1-\alpha_2|^2})+\nonumber\\
&+&\frac{1}{2}(1-e^{-\frac{1-|\lambda_1|^2}{2-3|\lambda_1|^2}|\alpha_1-\alpha_2|^2}).
\end{eqnarray}
The optimization of $|\lambda_1|$ for any value of $|\alpha_1-\alpha_2|$ yields $|\lambda_1|^2=1/3$, which corresponds to a performance of the  setup proposed in Paper I and hence concludes the proof of the optimality of that setup under the considered constraints.

\section{Evaluation of Gaussian type of integrals}
\label{appintegral}
As we have seen the following type of integrals
\begin{eqnarray}
I_m=\frac{1}{(2\pi \sigma^2)^m}\int_{\complex^{m}}d\alpha_1\ldots d\alpha_m e^{-\sum_{i=1}^m \frac{|\alpha_i|^2}{2\sigma^2}-\frac{a}{b}|x+\sum_{i=1}^{m}\alpha_i|^2}
\nonumber
\end{eqnarray}
emerge often in our calculation for the noise model. These integrals can be evaluated recursively using the relation 
\begin{eqnarray}
\frac{1}{(2\pi \sigma^2)}\int_{\complex}d\alpha\ e^{-\frac{|\alpha|^2}{2\sigma^2}-\frac{a}{b}|x+\alpha|^2}=\frac{b}{b+2a\sigma^2}e^{-\frac{a}{b+2a\sigma^2}|x|^2}
\nonumber\\ \label{step}
\end{eqnarray}
we are going to derive now.
Left hand side (LHS) of Eq. (\ref{step}) can be rewritten using the following modification of the rectangular identity
\begin{eqnarray}
&k&|\beta-\alpha_1|^2+l |\beta-\alpha_2|^2= \\
&=&\Big|\sqrt{k+l}\beta- \frac{k\alpha_1+l \alpha_2}{\sqrt{k+l}}\Big|^2+\frac{kl}{k+l}|\alpha_1-\alpha_2|^2\nonumber
\end{eqnarray}
as
\begin{eqnarray}
LHS&=&\frac{e^{-\frac{a}{b+2a\sigma^2}|x|^2}}{(2\pi \sigma^2)}\int_{\complex}d\alpha\ e^{-\Big|\sqrt{\frac{1}{2\sigma^2}+\frac{a}{b}}\alpha - \frac{2b\sigma^2 }{b+2a\sigma^2}x\Big|^2}\nonumber\\
&=&\frac{e^{-\frac{a}{b+2a\sigma^2}|x|^2}}{(2\pi \sigma^2)}\int_{\complex}d\alpha\ e^{-\Big|\sqrt{\frac{1}{2\sigma^2}+\frac{a}{b}}\alpha\Big|^2} \nonumber\\
&=&\frac{b}{b+2a\sigma^2}e^{-\frac{a}{b+2a\sigma^2}|x|^2}\frac{1}{2\pi \sigma'^2}\int_{\complex}d\alpha\ e^{-\frac{|\alpha|^2}{2\sigma'^2}} \nonumber\\
&=&\frac{b}{b+2a\sigma^2}e^{-\frac{a}{b+2a\sigma^2}|x|^2},
\end{eqnarray}
where we have used the fact that we are integrating over whole complex plane. As a consequence, a constant shift of argument does not matter and the Gaussian distribution is normalized to unity. Hence we have proved Eq. (\ref{step}), which we can be rewritten as
\begin{eqnarray}
I_m(a,b)=\frac{b}{b+2a\sigma^2}I_{m-1}(a,b+2a\sigma^2).\nonumber
\end{eqnarray} From this recursive rule it follows that
\begin{eqnarray}
I_m(a,b)=\frac{b}{b+2a\sigma^2 m}e^{-\frac{a}{b+2ma\sigma^2}|x|^2},
\end{eqnarray}
which is the result we wanted to obtain.

\end{document}